\documentclass[dvips]{article}
\usepackage{supertabular,lscape,epsfig}
\usepackage{rotating}
\usepackage{amssymb}
\usepackage{amsmath}
\DeclareSymbolFont{ppa}{OT1}{ppl}{m}{it}
\DeclareMathSymbol{\vv}{\mathalpha}{ppa}{'166}

\DeclareSymbolFont{ppa}{OT1}{ppl}{m}{it}
\DeclareMathSymbol{\vv}{\mathalpha}{ppa}{'166}

\begin{document}

\newcommand{\dd}{\,{\rm d}}
\newcommand{\ie}{{\it i.e.},\,}
\newcommand{\etal}{{\it et al.\ }}
\newcommand{\eg}{{\it e.g.},\,}
\newcommand{\cf}{{\it cf.\ }}
\newcommand{\vs}{{\it vs.\ }}
\newcommand{\zdot}{\makebox[0pt][l]{.}}
\newcommand{\up}[1]{\ifmmode^{\rm #1}\else$^{\rm #1}$\fi}
\newcommand{\dn}[1]{\ifmmode_{\rm #1}\else$_{\rm #1}$\fi}
\newcommand{\upd}{\up{d}}
\newcommand{\uph}{\up{h}}
\newcommand{\upm}{\up{m}}
\newcommand{\ups}{\up{s}}
\newcommand{\arcd}{\ifmmode^{\circ}\else$^{\circ}$\fi}
\newcommand{\arcm}{\ifmmode{'}\else$'$\fi}
\newcommand{\arcs}{\ifmmode{''}\else$''$\fi}
\newcommand{\MS}{{\rm M}\ifmmode_{\odot}\else$_{\odot}$\fi}
\newcommand{\RS}{{\rm R}\ifmmode_{\odot}\else$_{\odot}$\fi}
\newcommand{\LS}{{\rm L}\ifmmode_{\odot}\else$_{\odot}$\fi}

\newcommand{\Abstract}[2]{{\footnotesize\begin{center}ABSTRACT\end{center}
\vspace{1mm}\par#1\par
\noindent
{~}{\it #2}}}

\newcommand{\TabCap}[2]{\begin{center}\parbox[t]{#1}{\begin{center}
  \small {\spaceskip 2pt plus 1pt minus 1pt T a b l e}   
  \refstepcounter{table}\thetable \\[2mm]
  \footnotesize #2 \end{center}}\end{center}}

\newcommand{\TableSep}[2]{\begin{table}[p]\vspace{#1}
\TabCap{#2}\end{table}}

\newcommand{\FigCap}[1]{\footnotesize\par\noindent Fig.\  %
  \refstepcounter{figure}\thefigure. #1\par}

\newcommand{\TableFont}{\footnotesize}
\newcommand{\TableFontIt}{\ttit}
\newcommand{\SetTableFont}[1]{\renewcommand{\TableFont}{#1}}

\newcommand{\MakeTable}[4]{\begin{table}[t]\TabCap{#2}{#3}
  \begin{center} \TableFont \begin{tabular}{#1} #4 
  \end{tabular}\end{center}\end{table}}

\newcommand{\MakeTableSep}[4]{\begin{table}[p]\TabCap{#2}{#3}
  \begin{center} \TableFont \begin{tabular}{#1} #4
  \end{tabular}\end{center}\end{table}}
\newcommand{\TabCapp}[2]{\begin{center}\parbox[t]{#1}{\centerline{
  \small {\spaceskip 2pt plus 1pt minus 1pt T a b l e}
  \refstepcounter{table}\thetable}
  \vskip2mm
  \centerline{\footnotesize #2}}
  \vskip3mm
\end{center}}

\newcommand{\MakeTableSepp}[4]{\begin{table}[p]\TabCapp{#2}{#3}\vspace*{-.7cm}
  \begin{center} \TableFont \begin{tabular}{#1} #4 
  \end{tabular}\end{center}\end{table}}

\newfont{\bb}{ptmbi8t at 12pt}
\newfont{\bbb}{cmbxti10}
\newfont{\bbbb}{cmbxti10 at 9pt}
\newcommand{\uprule}{\rule{0pt}{2.5ex}}
\newcommand{\douprule}{\rule[-2ex]{0pt}{4.5ex}}
\newcommand{\dorule}{\rule[-2ex]{0pt}{2ex}}
\def\thefootnote{\fnsymbol{footnote}}

\newenvironment{references}%
{
\footnotesize \frenchspacing
\renewcommand{\thesection}{}
\renewcommand{\in}{{\rm in }}
\renewcommand{\AA}{Astron.\ Astrophys.}
\newcommand{\AAS}{Astron.~Astrophys.~Suppl.~Ser.}
\newcommand{\ApJ}{Astrophys.\ J.}
\newcommand{\ApJS}{Astrophys.\ J.~Suppl.~Ser.}
\newcommand{\ApJL}{Astrophys.\ J.~Letters}
\newcommand{\AJ}{Astron.\ J.}
\newcommand{\IBVS}{IBVS}
\newcommand{\PASP}{P.A.S.P.}
\newcommand{\Acta}{Acta Astron.}
\newcommand{\MNRAS}{MNRAS}
\renewcommand{\and}{{\rm and }}
\section{{\rm REFERENCES}}
\sloppy \hyphenpenalty10000
\begin{list}{}{\leftmargin1cm\listparindent-1cm
\itemindent\listparindent\parsep0pt\itemsep0pt}}%
{\end{list}\vspace{2mm}}

\def\TYLDA{~}
\newlength{\DW}
\settowidth{\DW}{0}
\newcommand{\dw}{\hspace{\DW}}

\newcommand{\refitem}[5]{\item[]{#1} #2%
\def\REFARG{#3}\ifx\REFARG\TYLDA\else, {\it#3}\fi
\def\REFARG{#4}\ifx\REFARG\TYLDA\else, {\bf#4}\fi
\def\REFARG{#5}\ifx\REFARG\TYLDA\else, {#5}\fi.}

\newcommand{\Section}[1]{\section{\hskip-6mm.\hskip3mm#1}}
\newcommand{\Subsection}[1]{\subsection{#1}}
\newcommand{\Acknow}[1]{\par\vspace{5mm}{\bf Acknowledgements.} #1}
\pagestyle{myheadings}

\newcommand{\xrule}{\rule{0pt}{2.5ex}}
\newcommand{\xxrule}{\rule[-1.8ex]{0pt}{4.5ex}}
\def\thefootnote{\fnsymbol{footnote}}

\begin{center}
{\Large\bf Searching for flickering variability in several\\ 
\vskip3pt
symbiotic stars and related objects:\\ 
\vskip6pt
BX~Mon, V471~Per, RS~Oph, V627~Cas, CI~Cam\\ 
\vskip6pt
V886~Her, Z~And, T~CrB, MWC~560, V407~Cyg\footnote{The original 
differential light curves $m_{var}$, $m_{comp}$ are available in anonymous ftp  at {\it sirius.astrouw.edu.pl}.}}
\vskip1.2cm
{\bf M.~~G~r~o~m~a~d~z~k~i$^{1}$,~~M.~~M~i~k~o~{\l}~a~j~e~w~s~k~i$^{2}$,~~T.~~T~o~m~o~v$^{2}$,\\
I.~~B~e~l~l~a~s~-~V~e~l~i~d~i~s$^3$,~~A.~~D~a~p~e~r~g~o~l~a~s$^3$,~~and~~C.~~G~a~{\l}~a~n$^2$}
\vskip8mm
{$^1$Centrum Astronomiczne im. Miko{\l}aja Kopernika, Polska Akademia Nauk,
ul. Bartycka 18, PL-00-716 Warszawa, Poland\\
e-mail: {\small marg@camk.edu.pl}\\
$^2$Uniwersytet Miko{\l}aja Kopernika, Centrum Astronomii,
ul. Gagarina 11, PL-87-100 Toru{\'n}, Poland\\
e-mail: {\small mamiko,tomtom,cgalan@camk.edu.pl}\\
$^3$National Observatory Athens, Institute of Astronomy and Astrophysics,
P.O. Box 20048, GR-11810 Athens, Greece\\
e-mail: {\small ibellas,adaperg@camk.edu.pl}\\}

\end{center}

\vskip1.6cm

\Abstract{{\it UBVRI} photometry observations of 10 symbiotic stars and related objects
obtained in the period 2002-2003 are presented. Analysing differential light
curves we found rapid light variations with timescales of tens of minutes
and significant amplitudes in the well-known flickerers MWC~560, RS~Oph,
V407~Cyg and T CrB. MWC~560 and V407~Cyg demonstrate quasi periodic
oscillations (QPO) with similar amplitudes and timescales. Flickering and
unusual flare in V627~Cas as well as some indications of flickering presence
in BX~Mon are detected.  The existence of 29 minutes oscillations in Z And
with an amplitude $\sim$0.02 mag in U band is confirmed. Only one symbiotic
star, V471~Per, and both non symbiotic, CI~Cam and V886~Her seem to be
constant on flickering timescales. Nevertheless, small night to night
changes in the brightness of V886~Her were observed as well.}

\Section{Introduction}
Symbiotic stars are wide binary systems, which generally consist of an
active white dwarf and a red giant. The binary is immersed in a nebula
formed by the red giant's wind. A part of this matter is accreted by the
white dwarf. The radiation of the hot companion partly ionizes the nebula
giving rise to characteristic symbiotic spectrum (highly ionized emission
lines superimposed on a cool K-M continuum). Symbiotic binaries show very
complex behavior: high and low activity stages, outbursts, jet ejections,
eclipses, pulsations of the cool star, etc.

Among about 200 symbiotic stars known, only 8 present rapid variability:
RS~Oph, T~CrB, MWC~560, Z~And, V2116~Oph, CH~Cyg, RT~Cru, o~Cet (Belczy\'nski 
\etal 2000, and refs. therein). Recently, V407 Cyg also has been reported as
a rapid flickerer in the V light (Kolotilov \etal 2003). Most of them show typical
flickering i.e. stochastic light variations on timescales from few minutes
to more than an hour. The typical flickering amplitude in symbiotic stars is
in the range 0.1-0.8 mag. Only Z And presents low-amplitude, strictly
coherent variations with a period of about half an hour (Sokoloski \etal 1999).

Flickering appears in high  activity stage and/or during outbursts. On the
other hand, during the eclipse of the hot companion in CH~Cyg no flickering
has been noticed (Miko{\l}ajewski \etal 1990; Sokoloski \& Kenyon 2003). These facts indicate that the
flickering is caused by changes in the surroundings of the compact
component. Most of the mentioned 8 stars include a hot component with
luminosity $\sim$10-100 L$_\odot$. Exceptions are Z~And ( the hot component
luminosity achieves 1000 L$_\odot$) and V2116~Oph ( the hot component is a
neutron star).

To explain the flickering origin in low-luminosity symbiotic systems
 Miko{\l}a-jewski \etal (1996a) adopted the propeller-accretor model (Lipunov 1987).
According to this model the flickering results from the interaction of the
magnetic white dwarf magnetosphere with the red giant's wind. Recently,
three-dimensional MHD simulations for a disc in propeller regime and Bondi
accretion confirmed the appearance of quasi-periodic oscillations (QPO)
comparable with the observed time-scale (Romanova \etal 2003; Romanova \etal 2004a).
Coherent periodic variations are expected only for a disc in accretor regime
(Romanova \etal 2004b). This is probably the case of Z And (Sokoloski \& Bildsten 1999).
On the other hand, (Sokoloski \& Kenyon 2003) consider that in the case of CH Cyg the
flickering originates in the inner parts of the accretion disk.

In this paper we present and discuss the results from photometric monitoring
of 8 symbiotic stars (V407~Cyg, V471~Per, MWC~560, Z~And, BX~Mon, T~CrB,
V627~Cas, RS~Oph) and of two related objects: the proto-planetary nebula
V886~Her and a high mass X-ray Binary CI~Cam.

\Section{Observations and Data Reductions}
Present observations have been carried out with the 1.2-m telescope at Kryoneri
Astronomical Station of the National Observatory Athens, Greece. The
detector was a 516$\times$516 pixels, UV-coated SI-502 CCD Camera with read
noise 9.0 $e^{-}$ rms and gain 5.17 e$^{-}$\slash ADU.  The peltier-cooled
to $-40$ degrees CCD chip gives dark current 1.03 e$^{-}$\slash pix\slash s.
The field of view was relative small, about 2.5$\times$2.5 arcmin and
2$\times$2 binning mode was applied. A standard set of {\it UBVRI} filters with
response curves close to those listed in Bessell (1990) was used.

Typical exposure times were in {\it U}: 40-90$^s$, in {\it B}: 20-90$^s$, in {\it V}:
10-40$^s$, in {\it R}: 4-20$^s$ and in {\it I}: 1-10$^s$. The time separation between
the measurements in the same filter was in the range of 2.5 - 5 minutes.
Standard reduction was preformed with IRAF\footnote{IRAF is distributed by
the National Optical Astronomy Observatories, which are operated by the
Association of Universities for Research in Astronomy, Inc., under
cooperative agreement with the National Science Foundation.}. Sky-subtracted
instrumental magnitudes of target and comparison stars were obtained by
means of IRAF task PHOT.

Our observations were restricted by two factors: the weather and the CCD
camera small field of view. In the instrumental light curves, besides
quadratic trend, a few minima caused by clouds can be seen. Because of this
a differential photometry method has been implemented. 

The best would be to use the method proposed by Sokoloski \etal (2001). They
constructed differential light curves by means of artificial standard star,
which was made of few stars.  We cannot apply the same method because of the
small field of view of our camera. Instead, the magnitudes of the target
star and of only two brightest field stars were measured. Two differential
curves were used for each observation run of each target star. The first one
presents the instrumental magnitude difference between the fainter and the
brighter comparison stars (thereafter $m_{comp}$). The second one shows the
difference between the target and the brighter comparison instrumental
magnitude ($m_{var}$). The {\it UBVRI} magnitudes of the comparison stars used are
taken from Henden \& Munari (2000,2001) and ({\it ftp:\slash \slash
ftp.nofs.navy.mil\slash pub\slash outgoing\slash aah}) and are presented in
Table 1.

\Section{Flickering criteria}
The standard deviation of the $m_{var}$ light curve is marked as
$\sigma_{var}$, the standard deviation of the $m_{comp}$ light curve is
marked as $\sigma_{comp}$. In Table 2 and Table 3
calculated standard deviations and average magnitudes for both types of
differential light curves are presented. The number of counts even from the
weakest sources were larger than $\sim 10^{5}$, thus the photon noise (in
magnitudes $\sigma \sim N^{-1/2}$) and camera noise ($\sigma \sim N^{-1}$)
should be negligible.  The relation $\sigma \sim N^{-1/n}$ between our
brighter and fainter stars is very poor ($n\gtrsim8$). A very usefull
approximation for large differences between brightnesses gives the equation:

\begin{equation}
\frac{\sigma_{m_1}}{\sigma_{m_2}}=1-k(m_2-m_1),
\end{equation}

\noindent where $m_1$ and $m_2$ denote the magnitudes of two stars. Using
about 30 pairs of magnitudes in different filters we have found
$k=0.099\pm0.009$.  The value $k=0.1$ has been adopted for further
calculations. Because of the difference in magnitudes of our variables and
comparisons it is necessary to correct the observed $\sigma_{comp}$. The new
value $\sigma_{comp}^\prime$, calculated by the use of Eq.~2, corresponds to
the expected standard deviation of a constant star with the same brightness
as the target i.e. $\sigma_{comp}^\prime=\sigma_{comp}(m_{var})$:

\begin{equation}
\sigma_{comp}^\prime=\sigma_{comp}[1-0.1(m_{comp}-m_{var})].
\end{equation}

This means that if we have an object 5 magnitudes brighter than the
comparison a twice smaller standard deviation must be applied to check its
variability. The ratio $\sigma_{var}$\slash $\sigma_{comp}^\prime$ can
define the level of rapid variations of the target star in comparison to the
constant star.  They are listed in the last five columns of
Table~3. Now we can determine the natural criteria for the
existence of pronounced variability of the investigated objects:

\begin{itemize}
\item[(0${\sigma}$)]{ {$\sigma_{var}$\slash $\sigma_{comp}^\prime <$} 0.5 - the comparison star is too faint or non constant;}

\item[(1${\sigma}$)]{0.5 $\leq \sigma_{var}$\slash $\sigma_{comp}^\prime <$  1.5 - lack of significant
   variability on measured sensitivity level (possible periodic changes can be reveal);}

\item[(2${\sigma}$)]{1.5 $\leq \sigma_{var}$\slash $\sigma_{comp}^\prime<$ 2.5 - most probably flickering presents and it is easy to establish if variations are periodic or the changes in the different filters are correlated};

\item[(3${\sigma}$)]{2.5 $\leq \sigma_{var}$\slash $\sigma_{comp}^\prime$ - evident variability with full amplitude about 6 $\sigma_{var}$ and time scale between the observing run duration and the mean measurements time span.}
\end{itemize}

The expected flickering amplitude can be defined as:

\begin{equation}
A=6\sqrt{(\sigma_{var})^2-(\sigma_{comp}^\prime)^2}.
\end{equation}

\begin{table}[hpt]
\caption{List of the used comparison stars.}
\begin{center}
\begin{footnotesize}
\begin{tabular}{cccccc}\hline
USNO-B1.0  & U & B & V & R & I\\
&&&&&\\
{\bf BX Mon}&&&&&\\
0864-0141979  &  12.770  &  12.809  &  12.418  &  12.158  &  11.906\\
0864-0141995  &  13.523  &  13.487  &  12.924  &  12.572  &  12.241\\
&&&&&\\
{\bf V471 Per}&&&&&\\
1428-0063362  &  12.117  &  12.046  &  11.479  &  11.148  &  10.794\\
1428-0063398  &  14.105  &  12.495  &  11.030  &  10.273  &  9.597\\
&&&&&\\
{\bf RS Oph}&&&&&\\
0833-0368817  &  17.054  &  15.969  &  14.415  &  13.528  &  12.645\\
0833-0368883  &  17.527  &  16.826  &  15.551  &  14.837  &  14.110\\
&&&&&\\
{\bf V627 Cas}&&&&&\\
1488-0371126  &  12.831  &  12.733  &  12.103  &  11.746  &  11.390\\
1487-0368382  &  12.963  &  12.844  &  12.323  &  12.023  &  11.700\\
&&&&&\\
{\bf CI Cam}&&&&&\\
1460-0131981  &  15.961  &  15.332  &  14.141  &  13.399  &  12.608\\
1460-0132003  &  17.025  &  16.539  &  15.680  &  15.168  &  14.607\\
&&&&&\\
{\bf V886 Her}&&&&&\\
1141-0280206 & - & 13.8$^\mathrm a$ & - & 13.1$^\mathrm b$ & 12.86$^\mathrm c$ \\
1141-0280174 & - & 14.2$^\mathrm a$ & - & 13.3$^\mathrm b$ & 13.28$^\mathrm c$ \\
&&&&&\\
{\bf Z And}&&&&&\\
1387-0498636  &  14.820  &  14.766  &  14.205  &  13.868  &  13.516\\
1387-0498647  &  15.237  &  15.155  &  14.475  &  14.042  &  13.606\\
&&&&&\\
{\bf T CrB}&&&&&\\
1158-0231034  &  11.946  &  11.779  &  11.166  &  10.766  &  10.452\\
&&&&&\\
{\bf MWC 560}&&&&&\\
0822-0179335  &  14.358  &  12.509  &  10.784  &  9.778  &  8.720\\
0822-0179227  &  15.456  &  14.035  &  12.668  &  11.925  &  11.267\\
&&&&&\\
{\bf V407 Cyg}&&&&&\\
1357-0407034  &  13.688  &  13.582  &  12.908  &  12.522  &  12.141\\
1357-0407056  &  12.969  &  12.857  &  12.233  &  11.885  &  11.541\\
\hline
\end{tabular}
\end{footnotesize}
\end{center}
\label{tab:std}
\begin{footnotesize}
\begin{list}{}{}
\item[$^{\mathrm a}$] B magnitude from USNO-A2.0 catalog;
\item[$^{\mathrm b}$] R magnitude from USNO-A2.0 catalog;
\item[$^{\mathrm c}$] I magnitude from USNO-B1.0 catalog.
\end{list}
\end{footnotesize}
\end{table}

\begin{sidewaystable}[hpt]
\caption{Averages and standard deviations of $m_{var}$ differential light curves.}
\begin{center}
\begin{scriptsize}
\begin{tabular}{cccccccccccccc}
\hline
Object\slash Data & JD & Duration [h] & Number & U$_{var}$ & $\sigma_{var}^U$  & 
B$_{var}$ & $\sigma_{var}^B$ & V$_{var}$ & $\sigma_{var}^V$ & R$_{var}$ &
$\sigma_{var}^R$ & I$_{var}$ & $\sigma_{var}^I$\\ &&&&&&&&&&&&&\\ 
{\bf BX Mon}&&&&&&&&&&&&&\\
16\slash03\slash2002 & 2452350 & 5.8 & 96 & -0.484 & 0.030 & -0.252 & 0.016 & -0.857 & 0.011 & -1.756 & 0.008 & -3.168 & 0.007\\
&&&&&&&&&&&&&\\
{\bf V471 Per}&&&&&&&&&&&&&\\
24\slash11\slash2002 & 2452603 & 1.4 & 50 & 2.036 & 0.016 & 1.926 & 0.009 & 1.622 & 0.006 & 1.235 & 0.006 & 1.118 & 0.007\\
25\slash11\slash2002 & 2452604 & 0.7 & 15 & 2.029 & 0.028 & 1.939 & 0.014 & 1.643 & 0.038 & 1.250 & 0.027 & 1.132 & 0.030  \\
&&&&&&&&&&&&&\\
{\bf RS Oph}&&&&&&&&&&&&&\\
16\slash06\slash2002 & 2452442 & 0.8 & 20 & - & - & -3.445 & 0.057 & -3.254 & 0.033 & -3.339 & 0.024 & -3.491 & 0.022\\
27\slash08\slash2002 & 2452514 & 1.8 & 32 & - & - & -3.110 & 0.047 & -3.087 & 0.028 & -3.251 & 0.019 & -3.436 & 0.014\\
&&&&&&&&&&&&&\\
{\bf V627 Cas}&&&&&&&&&&&&&\\ 
28\slash08\slash2002 & 2452515 & 5.8 & 96 & - & - & 1.980$^\mathrm a$ & 0.016$^{\mathrm a}$ & 0.383 & 0.006 & -1.238 & 0.012 & -3.296 & 0.017\\ 
&&&&&&&&&&&&&\\
{\bf CI Cam}&&&&&&&&&&&&&\\
24\slash11\slash2002 & 2452603 & 1.5 & 25 & -3.872 & 0.052 & -2.919 & 0.016 & -2.567 & 0.012 & -2.723 & 0.015 & -2.853 & 0.015\\
25\slash11\slash2002 & 2452604 & 1.6 & 30 & -3.837 & 0.044 & -2.860 & 0.023 & -2.530 & 0.014 & -2.678 & 0.014 & -2.833 & 0.012\\
&&&&&&&&&&&&&\\
{\bf V886 Her}&&&&&&&&&&&&&\\
15\slash06\slash2002 & 2452441 & 1.9 & 30 & -3.580 & 0.014 & -2.392 & 0.007 & -1.739 & 0.007 & -1.370 & 0.007 & -0.914 & 0.006\\
16\slash06\slash2002 & 2452442 & 1.5 & 30 & -3.671 & 0.012 & -2.475 & 0.007 & -1.802 & 0.006 & -1.432 & 0.006 & -0.975 & 0.007\\
20\slash06\slash2002 & 2452446 & 2.2 & 25 & -3.670 & 0.013 & -2.472 & 0.008 & -1.786 & 0.004 & -1.401 & 0.005 & -0.922 & 0.005\\
&&&&&&&&&&&&&\\
{\bf Z And}&&&&&&&&&&&&&\\
15\slash06\slash2002 & 2452441 & 1.1 & 24 & -4.243 & 0.025 & -3.605 & 0.013 & -3.909 & 0.011 & -4.651 & 0.016 & - & -\\
29\slash09\slash2002 & 2452547 & 3.9 & 86 & -3.942 & 0.013 & -3.580 & 0.012 & -3.922 & 0.012 & -4.690 & 0.015 & - & -\\
&&&&&&&&&&&\\
{\bf T CrB}&&&&&&&&&&&\\
14\slash03\slash2002 & 2452348 & 2.5 & 41 & 0.535 & 0.074 & -0.446 & 0.016 & -1.149 & 0.016 & -1.863 & 0.012 & -3.232 & 0.010\\
15\slash03\slash2002 & 2452349 & 4.4 & 52 & 0.427 & 0.083 & -0.466 & 0.016 & -1.171 & 0.018 & -1.877 & 0.014 & -3.234 & 0.017\\
16\slash03\slash2002 & 2452350 & 2.4 & 17 & 0.218 & 0.058 & -0.508 & 0.032 & -1.183 & 0.022 & -1.893 & 0.022 & -3.247 & 0.021\\
&&&&&&&&&&&&&\\
{\bf MWC 560}&&&&&&&&&&&&&\\
15\slash03\slash2002 & 2452349 & 3.8 & 102 & - & - & -1.117 & 0.094 & -0.183 & 0.076 & 0.065 & 0.043 & -0.480 & 0.014\\
23\slash11\slash2002 & 2452602 & 2.1 & 55 & -3.540 & 0.146 & -1.362 & 0.123 & -0.300 & 0.120 & 0.039 & 0.079 & -0.342 & 0.021\\
24\slash11\slash2002 & 2452603 & 3.5 & 80 & -3.362 & 0.087 & -1.217 & 0.068 & -0.158 & 0.061 & 0.138 & 0.040 & -0.325 & 0.014\\
25\slash11\slash2002 & 2452604 & 3.3 & 84 & -3.499 & 0.102 & -1.433 & 0.086 & -0.363 & 0.074 & -0.000 & 0.054 & -0.351 & 0.020\\
29\slash11\slash2002 & 2452608 & 5   & 150 & -3.583 & 0.086 & -1.421 & 0.063 & -0.360 & 0.061 & -0.008 & 0.044 & -0.365 & 0.016\\
30\slash11\slash2002 & 2452609 & 3.3 & 100 & -3.568 & 0.102 & -1.375 & 0.082 & -0.326 & 0.075 & 0.027 & 0.054 & -0.353 & 0.026\\
&&&&&&&&&&&&&\\
{\bf V407 Cyg}&&&&&&&&&&&&&\\
29\slash11\slash2002 & 2452608 & 3.3 & 80 & 1.935 & 0.132 & 2.046 & 0.141 & 1.648 & 0.140 & 0.212 & 0.054 & -2.772 & 0.016\\
11\slash06\slash2003 & 2452802 & 4.7 & 100 & 1.059 & 0.088 & 0.844 & 0.064 & -0.076 & 0.042 & -1.631 & 0.020 & -4.143 & 0.019\\
12\slash06\slash2003 & 2452803 & 5.2 & 120 & 1.133 & 0.070 & 0.874 & 0.050 & -0.067 & 0.030 & -1.647 & 0.017 & -4.170 & 0.023\\
13\slash06\slash2003 & 2452804 & 5.3 & 130 & 1.109 & 0.088 & 0.853 & 0.058 & -0.084 & 0.034 & -1.667 & 0.016 & - & -\\
14\slash06\slash2003 & 2452805 & 5.8 & 150 & 1.097 & 0.065 & 0.817 & 0.046 & -0.107 & 0.028 & -1.665 & 0.013 & - & -\\
03\slash07\slash2003 & 2452824 & 5.3 & 135 & 1.390 & 0.043 & 0.853 & 0.019 & -0.284 & 0.010 & -1.882 & 0.008 & - & -\\
04\slash07\slash2003 & 2452825 & 4.6 & 120 & 1.355 & 0.034 & 0.824 & 0.017 & -0.318 & 0.012 & -1.910 & 0.010 & - & -\\
05\slash07\slash2003 & 2452826 & 4.2 & 110 & 1.286 & 0.044 & 0.781 & 0.017 & -0.358 & 0.012 & -1.938 & 0.009 & - & -\\
06\slash07\slash2003 & 2452827 & 3.4 & 90 & 1.216 & 0.022 & 0.733 & 0.013 & -0.395 & 0.008 & -1.964 & 0.008 & - & -\\
\hline
\end{tabular}
\end{scriptsize}
\end{center}
\begin{scriptsize}
\begin{list}{}{}
\item[$^{\mathrm a}$] The flare was excluded.
\end{list}
\end{scriptsize}
\end{sidewaystable}

\begin{sidewaystable}[hpt]
\caption{Averages and standard deviations of $m_{comp}$ differential light curves. In the last five columns the ratio $\sigma_{var}/\sigma_{comp}^\prime$ is shown.}
\begin{center}
\begin{scriptsize}
\begin{tabular}{cccccccccccccccc}
\hline
Object\slash Data & U$_{comp}$ & $\sigma_{comp}^U$  & B$_{comp}$ & $\sigma_{comp}^B$ & V$_{comp}$ & $\sigma_{comp}^V$ & R$_{comp}$ & $\sigma_{comp}^R$ & I$_{comp}$ & $\sigma_{comp}^I$
&\multicolumn{5}{c} {$\frac{\sigma_{var}/\sigma_{comp}}{1-0.1(m_{comp}-m_{var})} $}
\\ 
&&&&&&&&&&&U&B&V&R&I\\ 
{\bf BX Mon}&&&&&&&&&&&&&&&\\
16\slash03\slash2002 & 0.751 & 0.053 & 0.639 & 0.010 & 0.517 & 0.007 & 0.418 & 0.007 & 0.325 & 0.009 & 0.645 & 1.756 & 1.822 & 1.460 & 1.195 \\
&&&&&&&&&&&&&&&\\
{\bf V471 Per}&&&&&&&&&&&&&&&\\
24\slash11\slash2002 & 1.873 & 0.015 & 0.340 & 0.006 & -0.397 & 0.005 & -0.821 & 0.005 & -1.237 & 0.006 & 1.049 & 1.294 & 0.998 & 0.995 & 0.944 \\
25\slash11\slash2002 & 1.853 & 0.053 & 0.312 & 0.031 & -0.401 & 0.005 & -0.828 & 0.012 & -1.249 & 0.017 & 0.519 & 0.388 & 6.310 & 1.863 & 1.425 \\
&&&&&&&&&&&&&&&\\
{\bf RS Oph}&&&&&&&&&&&&&&&\\
16\slash06\slash2002 & - & - & 0.906 & 0.027 & 1.144 & 0.015 & 1.307 & 0.020 & 1.527 & 0.029 & - & 3.737 & 3.927 & 2.241 & 1.522 \\
27\slash08\slash2002 & - & - & 0.916 & 0.018 & 1.144 & 0.013 & 1.303 & 0.017 & 1.532 & 0.026 & - & 4.371 & 3.733 & 2.052 & 1.070 \\
&&&&&&&&&&&&&&&\\
{\bf V627 Cas}&&&&&&&&&&&&&&&\\
28\slash08\slash2002 & - & - & 0.137 & 0.005 & 0.214 & 0.004 & 0.264 & 0.010 & 0.315 & 0.016 & - & 2.702 & 1.475 & 1.412 & 1.663 \\
&&&&&&&&&&&&&&&\\
{\bf CI Cam}&&&&&&&&&&&&&&&\\
24\slash11\slash2002 & 1.072 & 0.155 & 1.229 & 0.038 & 1.509 & 0.025 & 1.733 & 0.017 & 2.112 & 0.114 & 0.664 & 0.720 & 0.810 & 1.592 & 0.261 \\
25\slash11\slash2002 & 1.082 & 0.100 & 1.252 & 0.028 & 1.533 & 0.019 & 1.761 & 0.027 & 2.065 & 0.048 & 0.866 & 1.395 & 1.241 & 0.932 & 0.490 \\
&&&&&&&&&&&&&&&\\
{\bf V886 Her}&&&&&&&&&&&&&&&\\
15\slash06\slash2002 & 1.109 & 0.024 & 0.861 & 0.010 & 0.677 & 0.005 & 0.571 & 0.007 & 0.396 & 0.005 & 1.098 & 1.037 & 1.846 & 1.241 & 1.381 \\
16\slash06\slash2002 & 1.107 & 0.024 & 0.847 & 0.010 & 0.672 & 0.009 & 0.569 & 0.008 & 0.396 & 0.006 & 0.957 & 1.048 & 0.886 & 0.938 & 1.352 \\
20\slash06\slash2002 & 1.112 & 0.033 & 0.843 & 0.007 & 0.663 & 0.006 & 0.544 & 0.005 & 0.379 & 0.007 & 0.755 & 1.710 & 0.883 & 1.241 & 0.821 \\
&&&&&&&&&&&&&&&\\
{\bf Z And}&&&&&&&&&&&&&&&\\
15\slash06\slash2002 & 0.428 & 0.038 & 0.379 & 0.027 & 0.275 & 0.013 & 0.182 & 0.024 & - & - & 1.235 & 0.800 & 1.455 & 1.290 & - \\
29\slash09\slash2002 & 0.427 & 0.024 & 0.385 & 0.012 & 0.274 & 0.010 & 0.190 & 0.018 & - & - & 0.962 & 1.657 & 2.068 & 1.628 & - \\
&&&&&&&&&&&&&&&\\
{\bf MWC 560}&&&&&&&&&&&&&&&\\
15\slash03\slash2002 & - & - & 1.603 & 0.015 & 1.814 & 0.010 & 2.147 & 0.013 & 2.608 & 0.014 & - & 8.608 & 9.496 & 4.177 & 1.447 \\
23\slash11\slash2002 & 1.305 & 0.075 & 1.570 & 0.009 & 1.864 & 0.006 & 2.109 & 0.006 & 2.598 & 0.012 & 3.776 & 19.336 & 25.523 & 16.604 & 2.479 \\
24\slash11\slash2002 & 1.229 & 0.074 & 1.563 & 0.016 & 1.851 & 0.016 & 2.106 & 0.021 & 2.586 & 0.018 & 2.174 & 5.886 & 4.771 & 2.371 & 1.097 \\
25\slash11\slash2002 & 1.374 & 0.109 & 1.617 & 0.016 & 1.871 & 0.010 & 2.129 & 0.010 & 2.602 & 0.012 & 1.825 & 7.734 & 9.529 & 6.861 & 2.365 \\
29\slash11\slash2002 & 1.279 & 0.055 & 1.592 & 0.010 & 1.884 & 0.009 & 2.134 & 0.011 & 2.596 & 0.013 & 3.043 & 9.017 & 8.739 & 5.090 & 1.749 \\
30\slash11\slash2002 & 1.326 & 0.046 & 1.640 & 0.017 & 1.898 & 0.016 & 2.148 & 0.021 & 2.613 & 0.022 & 4.343 & 6.906 & 6.028 & 3.264 & 1.680 \\
&&&&&&&&&&&&&&&\\
{\bf V407 Cyg}&&&&&&&&&&&&&&&\\
29\slash11\slash2002 & 0.725 & 0.017 & 0.720 & 0.006 & 0.687 & 0.007 & 0.629 & 0.007 & 0.583 & 0.016 & 6.927 & 20.749 & 18.247 & 8.050 & 1.505 \\
11\slash06\slash2003 & 0.723 & 0.012 & 0.717 & 0.007 & 0.682 & 0.006 & 0.649 & 0.010 & 0.593 & 0.015 & 7.095 & 9.028 & 7.574 & 2.591 & 2.406 \\
12\slash06\slash2003 & 0.724 & 0.015 & 0.716 & 0.010 & 0.678 & 0.010 & 0.645 & 0.013 & 0.590 & 0.026 & 4.483 & 4.922 & 3.241 & 1.697 & 1.688 \\
13\slash06\slash2003 & 0.726 & 0.018 & 0.716 & 0.013 & 0.685 & 0.011 & 0.641 & 0.012 & - & - & 4.709 & 4.401 & 3.348 & 1.733 & - \\
14\slash06\slash2003 & 0.723 & 0.016 & 0.720 & 0.010 & 0.688 & 0.008 & 0.642 & 0.009 & - & - & 3.916 & 4.556 & 3.802 & 1.878 & - \\
03\slash07\slash2003 & 0.727 & 0.011 & 0.723 & 0.006 & 0.688 & 0.005 & 0.640 & 0.007 & - & - & 3.666 & 3.126 & 2.215 & 1.528 & - \\
04\slash07\slash2003 & 0.732 & 0.011 & 0.724 & 0.008 & 0.693 & 0.009 & 0.647 & 0.011 & - & - & 2.910 & 2.104 & 1.483 & 1.221 & - \\
05\slash07\slash2003 & 0.729 & 0.010 & 0.722 & 0.006 & 0.685 & 0.006 & 0.648 & 0.009 & - & - & 4.168 & 2.817 & 2.233 & 1.348 & - \\
06\slash07\slash2003 & 0.727 & 0.011 & 0.720 & 0.007 & 0.688 & 0.008 & 0.643 & 0.007 & - & - & 1.907 & 1.855 & 1.121 & 1.545 & - \\
\hline
\end{tabular}
\end{scriptsize}
\end{center}
\end{sidewaystable}

\Section{Results}

\Subsection{{\it BX Mon}}

BX Mon is most probably an eclipsing symbiotic binary with 1401 days period
(Dumm \etal 1998). The cool component was classified as M5III
(M\"urset \& Schmid 1999). The hot component with mass $\sim$0.6 M$_{\odot}$ and
luminosity reaching $\sim$230 $L_{\odot}$ most probably is a shell flashed
white dwarf (Dumm \etal 1998).

First attempt to detect flickering was done by Dobrzycka \etal (1996) and it was
negative. Sokoloski \etal (2001) included this object in a list of stars suspected
to show flickering activity. BX Mon has been observed only one night.  It
was about 3 weeks after the mid-eclipse epoch (Dumm \etal 1998) and,
unfortunately, one month before a significant brightening of about 3
magnitudes in {\it U} and {\it B} (Miko{\l}ajewski \& Ga{\l}an, in preparation).  The
observed by us variations in {\it BVR} filters (Table~3) fulfil the
third of our criteria as defined in Section~3. Moreover, the differential
$m_{var}$ light curves in these filters (Fig.~1) show similar
shapes that indicate possible variability on timescales of 1-2 hours. Such
correlations are not observed in the corresponding $m_{comp}$ light curves.
Flickering variability in BX~Mon is possible, however new observations,
especially during outbursts are needed.

\Subsection{{\it V471 Per}}

V471 Per belongs to the small group of yellow symbiotic stars. Such systems
have cool companion of F, G, K or early R spectral type (Miko{\l}ajewska 1997).
The spectral type of the V471 Per cool component is G5II
(Belczy\'nski \etal 2000). The nature of its hot component is unknown.

An attempt to detect flickering in V471 Per was made by Sokoloski \etal (2001) and
the result was negative. Our observations of this star were carried out
during two nights.  During the first night the
$\sigma_{var}/\sigma_{comp}^\prime$ ratio is close to 1 in all filters and $m_
{var}$ light curves show no variability (Fig.~2).

The observations during the second night formaly show detection of
flickernig in the {\it V} filter: $\sigma_{var}/\sigma_{comp}^\prime$ = 6.3
(Table~3).  However, the observing run was very short (15 points in
32 minutes) and additionally the weather conditions were not good close to
the end of this run when the detected variations apeared (Fig.~2). 
No corelation between the light-curve changes in the
different filters were observed. Concluding, this detection is very
questionable.

\Subsection{{\it RS Oph}}

RS Oph is one of the four recurrent novae among the symbiotic stars
(Anupama \&  Miko{\l}ajewska 1999).  The cool component is a K7III star (M\"urset \& Schmid 1999)
while the hot one is a white dwarf with luminosity $\sim$100 $L_{\odot}$
(Snijders 1987). Nova-like outbursts observed in this system have been
reported (Allen 1984; Kenyon 1986; Sanduleak \& Stephenson 1973; Schaefer 2004).
Flickering in RS Oph has been known since fifties. Rapid light variation
with amplitude $\sim$0.3 mag in RS Oph were reported by Walker (1957),
Walker (1977), Bruch (1980,1992). Brightness variations on a
timescale of about 80 minutes was found by Dobrzycka \etal (1996).

We observed RS Oph during two nights. In both cases the duration of the
observations was comparable with the typical timescale of the flickering in
this system. A strong linear decrease is evident in the $m_{var}$
differential light curves (Fig.~3). Theirs slopes on 27 August
2002 were 0.16, 0.10, 0.08, 0.05 mag in {\it BVRI} filters respectively.
Despite the very short duration of our observations on 16 June the slopes
are even greater. However, the {\it I} band inversed slopes of $m_{var}$ and
$m_{comp}$ curves (Fig.~3) can be connected with the
second-order atmospheric extinction. The existence of short time variability
in {\it BV} is evident also in the sense of our criterion (3${\sigma}$).

\Subsection{{\it V627 Cas}}

V627 Cas has been initially classified as T Tau variable (Paupers \etal 1989),
whereas Kolotilov (1998) proposed its symbiotic nature. It was included in
the Belczy\'nski \etal (2000) list of stars suspected to be symbiotic. The
spectral type of the cool component is M2-4II (Kolotilov \etal 1991). Pulsations
with a period of $\sim$466 days and brightness changes from night to night,
indicating possible presence of flickering, were reported by Kolotilov \etal (1991,1996). 
It has been suggested that the V627 Cas hot companion activity is of
 similar nature as that of CH Cyg.

Kolotilov \etal (1991) also observed short-living flares with amplitude of about
0.5 magnitudes in {\it B}. A similar phenomenon was registered in our data. A
flickering-like flare with amplitude $\sim$0.1 mag and duration $\sim$30
minutes in {\it B} is evident on August 28 whereas in the other bands ({\it VRI}) no
such brightness changes were noticed (Fig.~4).  The ratio
$\sigma_{var}$\slash $\sigma_{comp}^\prime$ is about 2.7 in {\it B} (the flare
excluded) and $\sim$1.5 in the other bands (Table~3). The lack of
flares in the other filters can be caused by the significant cool component
contribution. Additionally, a weak linear trend with decreasing rate of
0.01 mag per hour is evident in the $m_{var}$ differential light curve in {\it B} .
Similar but smaller slopes in $m_{var}$ are visible in the other filters as
well (Fig.~4). Such linear trends can reflect night to night
brightness changes on time scales few times longer than the typical
flickering ones.

\Subsection{{\it CI Cam}}

CI Cam is a high mass X-ray binary which includes a B[e]II star as a mass
donor. The nature of the compact star is unclear.  However,
$L(2-25$keV$)>3\times10^{38}$ ergs s$^{-1}$ is typical for a binary with a
black hole component but, there are no mass estimations of the compact
object (Robinson \etal 2002). A strong outburst occurred in March 1998 and CI
Cam became the brightest object in X-rays on the whole sky. It was two times
brighter than Crab (Smith \etal 1999). Five days later
Hjellming \& Mioduszewski (1998a,b) observed appearance of radio
jets with velocity of 0.15 ${\it c}$.  Before the outburst,
Bergner \etal (1995) speculated about symbiotic nature of CI Cam.

Flickering in this system has been never detected. One negative attempt was
made by Belloni \etal (1999). Our observations show
$\sigma_{var}$\slash$\sigma_{comp}^\prime$ less than 1 for almost all filters (Table~3) 
i.e. lack of flickering with amplitude larger than 0.04-0.05 for
{\it U} and 0.01-0.02 for {\it BVRI} bands (Table~2).

\Subsection{{\it V886 Her}}

V886 Her = IRAS 18062+2410 was classified as a proto-planetary nebula and
its mass was estimated to 0.7$\pm$0.01 M$_{\odot}$ (Arkhipova \etal 1996).  The
star's spectral type changes relatively fast. From A5, as listed in the HDE
catalog, to B5I or B1-2II estimated by Arkhipova \etal (1996, 1999). V886 Her
is a Be star with strong H$\alpha$ (EW $\approx 20$ {\AA}) and other Balmer and
strong oxygen OI $\lambda$8446{\AA} emission lines. Arkhipova \etal (1999) found
that V886~Her is embedded in a dust envelope with a temperature about 220~K.

Brightness variations with an amplitude of 0.4 mag and timescales from a day
to a week has been observed in 1995 (Arkhipova \etal 1996). Gauba \& Parthasarathy (2003)
noticed changes in the UV (IUE) flux during three years.

Our observations of V886~Her cover three nights. We did not find clear
evidences of flickering but weak linear trends in {\it U}, {\it B} and {\it R} of 0.005-0.01
magnitudes per hour are present (Fig.~5). These trends can
explain the small changes of the mean brightness from night to night up to
0.09 magnitudes in {\it U}, 0.08 in {\it B} and about 0.06 in {\it VRI} that were observed as
well.

\Subsection{{\it Z And}}

Z And is a prototype of symbiotic star and is one of the most intensively
studied. The spectral type of the cool component is M4.5III
(M\"urset \& Schmid 1999) whereas the companion is a hot white dwarf ($T_{\rm
eff}\sim10^5$ K) with luminosity $\sim$500-2000 $L_{\odot}$
(Viotti \etal 1982; Kenyon \& Webbink 1984; Fern\'andez-Castro \etal 1988; M\"urset \etal 1991).
A stable nuclear
burning of accreted matter on the compact component surface is considered as
the most probable energy source. Sporadic changes in the accretion rate
cause irregular outbursts with an amplitude of about 2-3 mag
(Miko{\l}ajewska \& Kenyon 1992)

The orbital period of Z And is 759 days
(Formiggini \& Leibowitz 1994; Miko{\l}ajewska \& Kenyon 1996). The inclination angle estimated
on the base of polarimetric observations is 45$\pm$12 degree which gives for
the mass of the white dwarf 0.65$\pm$0.28 $M_{\odot}$ (Schmid \& Schild 1997).
Possible radio jets emission has been reported by (Brocksopp \etal 2004).

No evidence of classical, stochastic flickering has been noticed in several
sets of long-term monitoring of Z And
(Walker 1957; Belyakina 1965; Dobrzycka \etal 1996). For the first time, during the
1997 outburst, Sokoloski \& Bildsten (1999) reported the presence of strictly periodic
rapid light variability, shortly after the maximum (Fig.~6). 
Contrary, during the quiescence ($\sim$JD 24501000) no
rapid variations were presented.

Our observations cover two nights. In both cases  
$\sigma_{var}/\sigma_{comp}^\prime$ $\leq 1.5$ indicates no flickering. However,
the  detection level was about 0.011-0.025 magnitude (Table~3) i.e.
few times larger than the above mentioned amplitude of periodic
oscillations. Following Sokoloski \& Bildsten (1999), we tried to search for periodicity
in the pure instrumental data after quadratic trend (air mass changes)
removing. The Fourier analysis (Fig.~7) of such light curves
from the second night (when the weather conditions were pretty good) reveals
$\sim$29 minutes period in {\it U}, very close to that found by Sokoloski \& Bildsten (1999).
The folded light curve with this period (1781~sec) is shown in the bottom
panel of Fig.~7. The amplitude in {\it U} is about 0.02 magnitude.
Our observations were carried out just before and just after a short, deep
minimum during the 2000-2003 outburst (Fig.~6).
Skopal (2003) interpreted this minimum as an eclipse of the hot
component. The {\it UBV} brightnesses of Z~And at the moments of our observations
and the observations of Sokoloski \& Bildsten (1999) during the previous outburst are
similar(Fig.~6). They discovered coherent oscillations in {\it B}
filter with a period 28 min and an amplitude about 3 mmag. Our oscillations
in U filter show almost the same period and an amplitude about 15-20 mmag,
i.e. about 5-7 times larger than the previously observed in {\it B}.  Suggesting
that the rapid variations in {\it U} and {\it B} have the same origin, the amplitude
difference argues the higher contribution of the oscillations source towards
the short wavelengths.

\Subsection{{\it T CrB}}

T CrB is another recurrent nova among the symbiotic stars. Two outbursts in
this system were observed in 1866 and 1946 (e.g. Belczy\'nski \&
Miko{\l}ajewska 1998).  In quiescence the system reveals an ellipsoidal
variability with a period of $\sim$228 days. This fact suggests that the
cool component almost fills up its Roche lobe. The spectral type of the cool
star is M4III (M\"urset \& Schmid 1999). The white dwarf luminosity is $\sim$5-40
$L_{\odot}$ (Cassatella \etal 1982,1985; Selvelli \etal 1992). For a long time there was
a controversy concerning T CrB because the compact component mass
estimations exceeded the Chandrasekhar limit
(Kraft 1958; Paczy\'nski 1965; Kenyon \& Garcia 1986).
Belczy\'nski \& Miko{\l}ajewska (1998) lowered this mass to the range 1.2-1.4 M$_{\odot}$. Recently, 
Stanishev \etal (2004) analyzing H$_{\alpha}$ observations of T~CrB obtained during 
the last decade, found an orbital inclination $i$=67$^o$ and components mass 
 M$_{WD}$=1.37$\pm$0.13 M$_{\odot}$ and M$_{RG}$=1.12$\pm$0.23 M$_{\odot}$. 

Strong flickering in {\it U} with amplitude of 0.1-0.5 has been known in
T~CrB since long time
(Ianna 1964; Lawrence \etal 1967; Bianchini \& Middleditch 1976; Sokoloski \etal 2001; Zamanov \etal 2004)
In {\it B} and
{\it V} the flickering amplitude is smaller (Raikova \& Antov 1986; Hric \etal 1998). However,
observations with no flickering detection has been reported as well
(Bianchini \& Middleditch 1976; Oskanian 1983; Dobrzycka \etal 1996; Miko{\l}ajewski \etal 1997).

Because of the small field of view, during the three nights of our
monitoring, we observed T~CrB and the comparison star alternatively, on
different CCD frames. The brightness of the comparison star was interpolated
to the observation time of T~CrB and only $m_{var}$ differential light
curves were obtained. In these curves relatively high amplitude rapid
brightness variations and up to $\sim$0.2 mag night to night changes of the
mean brightness are evident only in {\it U}. Therefore, our results confirm the
cited above that, when presented in T~CrB, the flickering is best visible in
{\it U} band.

\Subsection{{\it MWC 560}}

MWC 560 is one of the most enigmatic symbiotic systems with spectacular jets
ejected along the line of sight (Tomov \etal 1990).  The spectral type of the
cool component is M4-6III 
(Thakar \& Wing 1992; Zhekov \etal 1996; M\"urset \& Schmid 1999). 
Depending
on the activity stage the luminosity of the white dwarf was estimated to
$\sim$100-1000 $L_{\odot}$ (Tomov \etal 1996; Miko{\l}ajewski \etal 1998).

The first detection of flickering has been reported by Bond \etal (1984).
Since 1990, when the systematic investigation of MWC~560 began, many
authors observed flickering with amplitudes in the range 0.2-0.7 mag
(Michalitsianos \etal 1993, Tomov \etal 1996; Dobrzycka \etal 1996;
Miko{\l}ajewski \etal 1998; Ishioka \etal 2001; Sokoloski \etal 2001).

Our observations of MWC 560 include six nights. The ratios
$\sigma_{var}$\slash$\sigma_{comp}^\prime$ together with roughly the same
light curves shape in all bands indicate presence of flickering in all
nights.  As an example the differential light curves for two nights are
presented in Fig~9.

Fourier analysis (Deeming 1975) of the MWC 560 differential light
curves do not show significant frequency peaks in the power spectra of
$m_{comp}$. Rather, a typical noise spectrum is seen. Contrary, the power
spectra of $m_{var}$ light curves are dominated by low frequency peaks
corresponding to periods in the range from 30 to 100 minutes. The shape of
the power spectra changes from night to night. An analysis of combined data
set (23, 24, 25, 29, and 30 November) do not show coherent variations.  The
power spectra are dominated by low frequency peaks with periodicity of
$\sim$2 hours, $\sim$70, $\sim$50 minutes (see Fig.~10). The
residuals of the light curves folded with these periods are presented in
Fig.~11. The period $\sim$70 minutes corresponds to the sharp
maximum of possible quasi-periods found by Tomov \etal (1996) in almost fifty
different runs of flickering monitoring. Regular, sinusoidal variations with
amplitude larger than $\pm3\sigma$ in {\it UBVR} (Fig.~11) demonstrates
only the mean light curve with $\sim$70 minutes period ($P=4241$~sec).
Moreover, Miko{\l}ajewski \etal (1996b), on data obtained also with the 1.2m Kryoneri
telescope during four nights of January 1993, estimated almost the same
frequency ($P=4225.3$~sec). It is rather difficult to argue about the
coherence of the variations with this period, however, the power spectra
suggest that such variations can appear and disappear. So, they should be
classified rather as a kind of quasi periodic oscillations (QPO).

\Subsection{{\it V407 Cyg}}

V407 Cyg is a symbiotic system which includes a low luminosity white dwarf
and a Mira with the longest known among the symbiotics pulsation period of
763 days. The luminosity of the hot component is $\sim$100-400 L$_{\odot}$
depending on the stage of activity (Kolotilov \etal 2003; Tatarnikova \etal 2003). The spectrum is
usually dominated by absorption features, originating in the cool component
atmosphere. Emission lines typical for a symbiotic system are sporadically
observed. For the first time a symbiotic spectrum was observed in 1994
(Kolotilov \etal 1998). Rapid photometric variability in V was detected in June
2002 (Kolotilov \etal 2003). The variations amplitude was 0.2 mag and Fourier
analysis showed two periods: 18 and 33 minutes.

The observations of V407~Cyg were secured during one night in November 2002,
four consecutive nights in June 2003 and four consecutive nights in July
2003. At that time the outburst which has began in 1998
(Kolotilov \etal 2003; Tatarnikova \etal 2003) still continued. After the third night we stopped
the observations in {\it I} band because Mira was too bright being in maximum.
Examples of differential light curves are presented in Fig.~12. The corrected ratio
$\sigma_{var}$\slash$\sigma_{comp}^\prime$ indicates the permanent presence
of flickering in {\it UBV} and far from the Mira maximum also in {\it R}.

The Fourier analysis (Deeming 1975) of V407 Cyg differential light
curves shows results similar to those of MWC 560. The $m_{var}$ light curves
power spectra in {\it UBVR} are very similar and are dominated by low frequency
peaks corresponding to periods in the range from 30 to about 100 minutes.
Contrary, the power spectra of $m_{var}$ in {\it I} do not show significant
frequency peaks, suggesting that the {\it I} flux is dominated by Mira and there
is no evident flickering. Two data sets, each combined by the observations
in four consecutive nights (June and July 2003) were analyzed as well. The
power spectra of the combined data are dominated by low frequency peaks
corresponding to periods of $\sim$2 hours, $\sim$72 and $\sim$45 minutes
(Fig.~13). As it can be seen in Fig.~13
significant night to night variations in the peaks are evident. Obviously,
the rapid variations in the brightness of V407~Cyg are not coherent. Most
probably these are QPO like the observed in MWC~560.

\Section{Concluding remarks}

In this paper we present results of one of the most extensive surveys
searching for flickering in symbiotic stars and related objects. Two earlier
surveys had been published by Dobrzycka \etal (1996) and Sokoloski \etal (2001). Most
of the previous flickering observations were made in one band only, usually
in {\it U} or {\it B}. We observed in five filters ({\it UBVRI}) allowing us to obtain more
detailed information about the contribution of the flickering source to the
symbiotic spectrum. Our results are summarized in Table~4, where the
flickering amplitudes are calculated using Equation (3) and assuming
corrected $\sigma_{comp}^\prime$ as their error.

We found that the flickering is presented in almost all bands in the
symbiotic systems MWC~560, RS~Oph and V407~Cyg. This proves that here the
contribution of the hot component to each particular band is sufficient.
Another symbiotic star with similar flickering characteristics is CH~Cyg
(Miko{\l}ajewski \etal 1990). The variations timescales and amplitudes in these four
systems show close resemblance and indicate a stochastic nature. In
particular nights, quasi-periodic oscillations with periods from several
minutes to more than one hour can dominate the light curve.  The QPO are not
coherent from night to night and over longer periods, and vary in amplitude
and phase.  Typical properties of these four stars are a low-luminosity hot
component and permanent or occasional jets ejection. The common
observational characteristics of the flickering and the physical properties
of these objects point to a similar flickering generation mechanism.

It seems that T~CrB should also be included in this group of stars, but its
low activity during our observation run allowed to detect flickering in {\it U}
band only. The symbiotic star V627~Cas, seems also to contain a
low-luminosity hot component. At least two types of variability are evident
in our observations: short-living flickering-like flares and night to night
changes. We cannot exclude the existence of flickering in BX~Mon.  The
differential light curves in {\it U} and {\it R} show almost the same shapes as in {\it BV},
where $\sigma_{var}/\sigma_{comp}^\prime$ are greater
than 1.5, satisfying our criterion ($2\sigma$).

Unlike the above stars, Z~And is a classical symbiotic system with a high
luminosity hot component during outbursts. Rapid variability was undoubtedly
observed in this system by Sokoloski \& Bildsten (1999), but only in {\it B}. It showed low
amplitude (0.002-0.005 mag) and strictly periodic variations, and probably
lack of stochastic changes. This result was interpreted as caused by the
rotation of the magnetic white dwarf component. There are not known coherent
periodicities in other symbiotic systems.  We found the same periodicity
during another outburst and with another filter. The variations here have
amplitude of about 0.02 magnitudes in {\it U}, i.e. several times larger than
previously observed in {\it B}.

\begin{table}[hpt]
\caption{Flickering amplitudes in different filters estimated according Eq. 3. 
As errors the $\sigma_{comp}^\prime$ values are assumed. The amplitudes derived from 
the data satisfying our criterion (1 $\sigma$) are shown  in brackets, because they are 
probably not real.}
\begin{center}
\begin{scriptsize}
\begin{tabular}{ccccccc}\hline
Object\slash Data & $A_U$ & $A_B$ & $A_V$ & $A_R$ & $A_I$ \\
&&&&&\\
{\bf BX Mon}&&&&&\\
16\slash03\slash2002 &  -    & 0.079$\pm$0.009 & 0.055$\pm$0.006 & (0.035$\pm$0.005) & (0.023$\pm$0.006) \\
&&&&&\\
{\bf V471 Per}&&&&&\\
24\slash11\slash2002 & (0.029$\pm$0.015) & (0.034$\pm$0.007) &  -    &  -    &  -    \\
25\slash11\slash2002 &  -    &  -    & 0.225$\pm$0.006: & 0.137$\pm$0.014: & (0.128$\pm$0.021) \\
&&&&&\\
{\bf RS Oph}&&&&&\\
16\slash06\slash2002 &  -    & 0.330$\pm$0.015 & 0.191$\pm$0.008 & 0.129$\pm$0.011 & 0.100$\pm$0.014 \\
27\slash08\slash2002 &  -    & 0.275$\pm$0.011 & 0.162$\pm$0.007 & 0.100$\pm$0.009 & (0.030$\pm$0.013) \\
&&&&&\\
{\bf V627 Cas}&&&&&\\
28\slash08\slash2002 &  -    & 0.089$\pm$0.006 & (0.026$\pm$0.004) & (0.051$\pm$0.008) & 0.081$\pm$0.010 \\
&&&&&\\
{\bf CI Cam}&&&&&\\
24\slash11\slash2002 &  -    &  -    &  -    & 0.070$\pm$0.009: &  -    \\
25\slash11\slash2002 &  -    & (0.096$\pm$0.016) & (0.050$\pm$0.011) &  -    &  -    \\
&&&&&\\
{\bf V886 Her}&&&&&\\
15\slash06\slash2002 & (0.035$\pm$0.013) & (0.011$\pm$0.007) & 0.035$\pm$0.004 & (0.025$\pm$0.006) & (0.025$\pm$0.004) \\
16\slash06\slash2002 &  -    & (0.013$\pm$0.007) &  -    &  -    & (0.028$\pm$0.005) \\
20\slash06\slash2002 &  -    & 0.039$\pm$0.005 &  -    & (0.018$\pm$0.004) &  -    \\
&&&&&\\
{\bf Z And}&&&&&\\
15\slash06\slash2002 & (0.088$\pm$0.020) &  -    & (0.048$\pm$0.008) & (0.061$\pm$0.012) &  -    \\
29\slash09\slash2002 & 0.020$\pm$0.003$^\mathrm a$  & 0.057$\pm$0.007: & 0.063$\pm$0.006: & 0.071$\pm$0.009: &  -    \\
&&&&&\\
{\bf T CrB}&&&&&\\
15\slash06\slash2002 & 0.32$\pm$0.05$^\mathrm b$ &  -    &  -    &  -    &  -    \\
16\slash06\slash2002 & 0.39$\pm$0.05$^\mathrm b$ &  -    &  -    &  -    &  -    \\
20\slash06\slash2002 & 0.18$\pm$0.05$^\mathrm b$ &  -    &  -    &  -    &  -    \\
&&&&&\\
{\bf MWC 560}&&&&&\\
15\slash03\slash2002 &  -    & 0.560$\pm$0.011 & 0.453$\pm$0.008 & 0.250$\pm$0.010 & (0.061$\pm$0.010) \\
23\slash11\slash2002 & 0.845$\pm$0.039 & 0.737$\pm$0.006 & 0.719$\pm$0.005 & 0.473$\pm$0.005 & 0.115$\pm$0.008 \\ 
24\slash11\slash2002 & 0.463$\pm$0.040 & 0.402$\pm$0.012 & 0.358$\pm$0.013 & 0.218$\pm$0.017 & (0.035$\pm$0.013) \\
25\slash11\slash2002 & 0.512$\pm$0.056 & 0.512$\pm$0.011 & 0.442$\pm$0.008 & 0.321$\pm$0.008 & 0.109$\pm$0.008 \\
29\slash11\slash2002 & 0.487$\pm$0.028 & 0.376$\pm$0.007 & 0.364$\pm$0.007 & 0.259$\pm$0.009 & 0.079$\pm$0.009 \\
30\slash11\slash2002 & 0.596$\pm$0.023 & 0.487$\pm$0.012 & 0.444$\pm$0.012 & 0.308$\pm$0.017 & 0.125$\pm$0.015 \\
&&&&&\\
{\bf V407 Cyg}&&&&&\\
29\slash11\slash2002 & 0.784$\pm$0.019 & 0.845$\pm$0.007 & 0.839$\pm$0.008 & 0.321$\pm$0.007 & 0.072$\pm$0.011 \\
11\slash06\slash2003 & 0.523$\pm$0.012 & 0.382$\pm$0.007 & 0.250$\pm$0.006 & 0.111$\pm$0.008 & 0.104$\pm$0.008 \\
12\slash06\slash2003 & 0.409$\pm$0.016 & 0.294$\pm$0.010 & 0.171$\pm$0.009 & 0.082$\pm$0.010 & 0.111$\pm$0.014 \\
13\slash06\slash2003 & 0.516$\pm$0.019 & 0.339$\pm$0.013 & 0.195$\pm$0.010 & 0.078$\pm$0.009 &  -    \\
14\slash06\slash2003 & 0.377$\pm$0.017 & 0.269$\pm$0.010 & 0.162$\pm$0.007 & 0.066$\pm$0.007 &  -    \\
03\slash07\slash2003 & 0.248$\pm$0.012 & 0.108$\pm$0.006 & 0.054$\pm$0.005 & 0.036$\pm$0.005 &  -    \\
04\slash07\slash2003 & 0.192$\pm$0.012 & 0.090$\pm$0.008 & (0.053$\pm$0.008) & (0.034$\pm$0.008) &  -    \\
05\slash07\slash2003 & 0.256$\pm$0.011 & 0.095$\pm$0.006 & 0.064$\pm$0.005 & (0.036$\pm$0.007) &  -    \\
06\slash07\slash2003 & 0.112$\pm$0.012 & 0.066$\pm$0.007 & (0.022$\pm$0.007) & 0.037$\pm$0.005 &  -    \\
\hline
\end{tabular}
\end{scriptsize}
\end{center}
\label{tab:amp}
\begin{scriptsize}
\begin{list}{}{}
\item[$^{\mathrm a}$] amplitude defined as the difference between maximum and minimum of the
   mean light curve with period P=1781s (Fig.~7).
\item[$^{\mathrm b}$] $\sigma^{\prime}_{comp}=0.05$ was adopted.
\end{list}
\end{scriptsize}
\end{table}

Among the observed symbiotic stars only V471~Per seems to be constant on
short time scales. We also did not found rapid variability in CI~Cam and
V886~Her, the two not symbiotic objects among our targets. However, the
central star of the protoplanetary nebula, V886~Her, shows different
magnitudes from night to night in {\it UBVRI} indicating a long-term variability
with amplitude of about 0.1 magnitude.

\vspace{5mm}
\Acknow{We would like to thank Prof. Joanna Miko{\l}ajewska for useful comments.
This work was partly supported by the Polish State Committee for Scientific
Research grants Nos. 1P03D 017 27, 5P03D 003 20. We would like to thank to
many our colleagues and students, especially to Ariel Majcher, Sylwia
Frackowiak, Karolina Wojtkowska, Jakub Janowski, Darek Graczyk and others,
for their help in assembling of the long-term light curve of Z~And at Torun
Observatory.}

\newpage
\begin{figure}
\includegraphics[width=10cm]{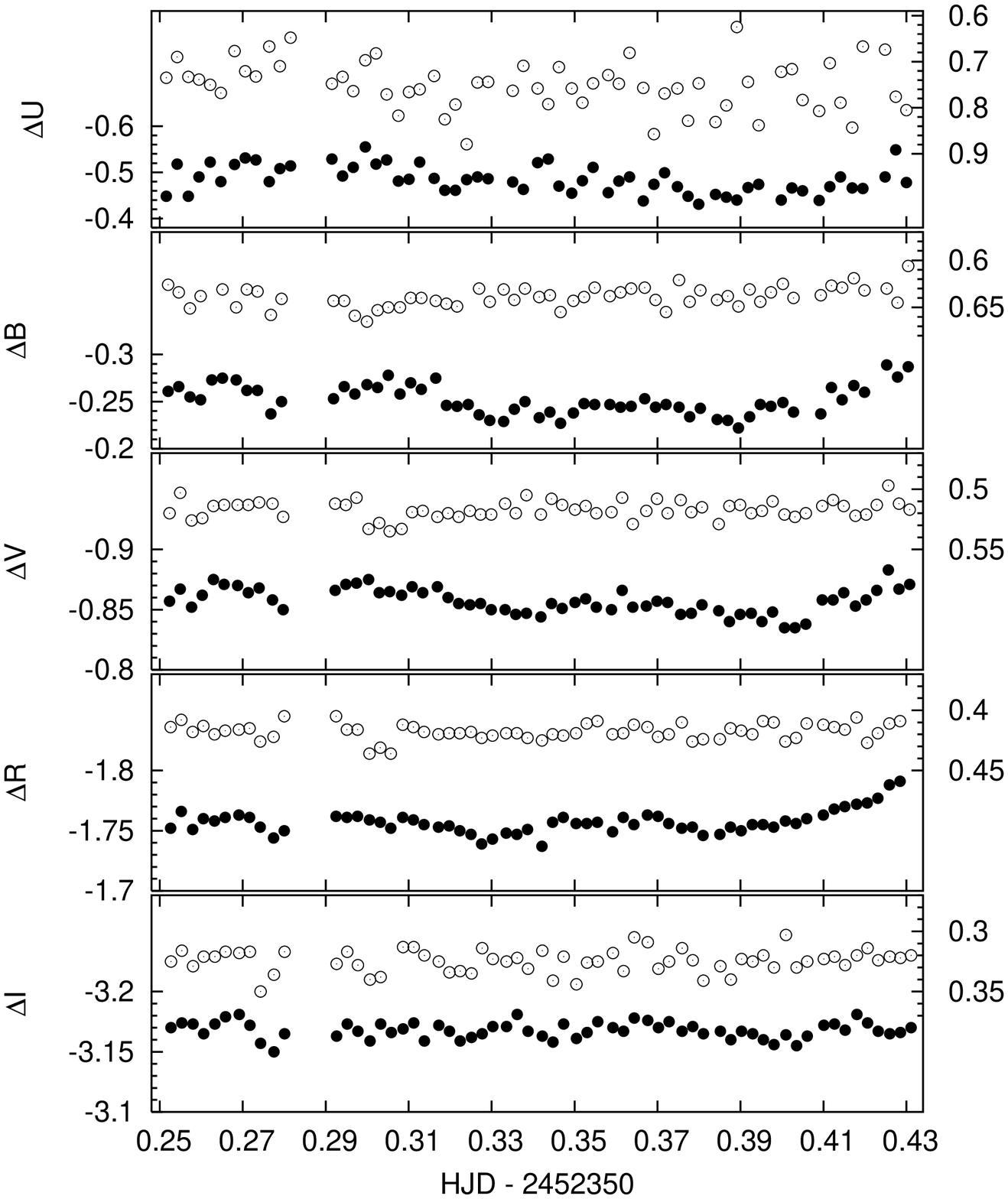}
\caption{Differential light curves $m_{var}$ (dots) and  $m_{comp}$ 
(open circles) for BX~Mon on 16 March 2002. The similar behaviour of the $m_{var}$ light
curves on timescales of 1-2 hours is evident in {\it BVR} (and possibly in {\it U}) filters.}
\end{figure}

\newpage
\begin{figure}
\includegraphics[width=10cm]{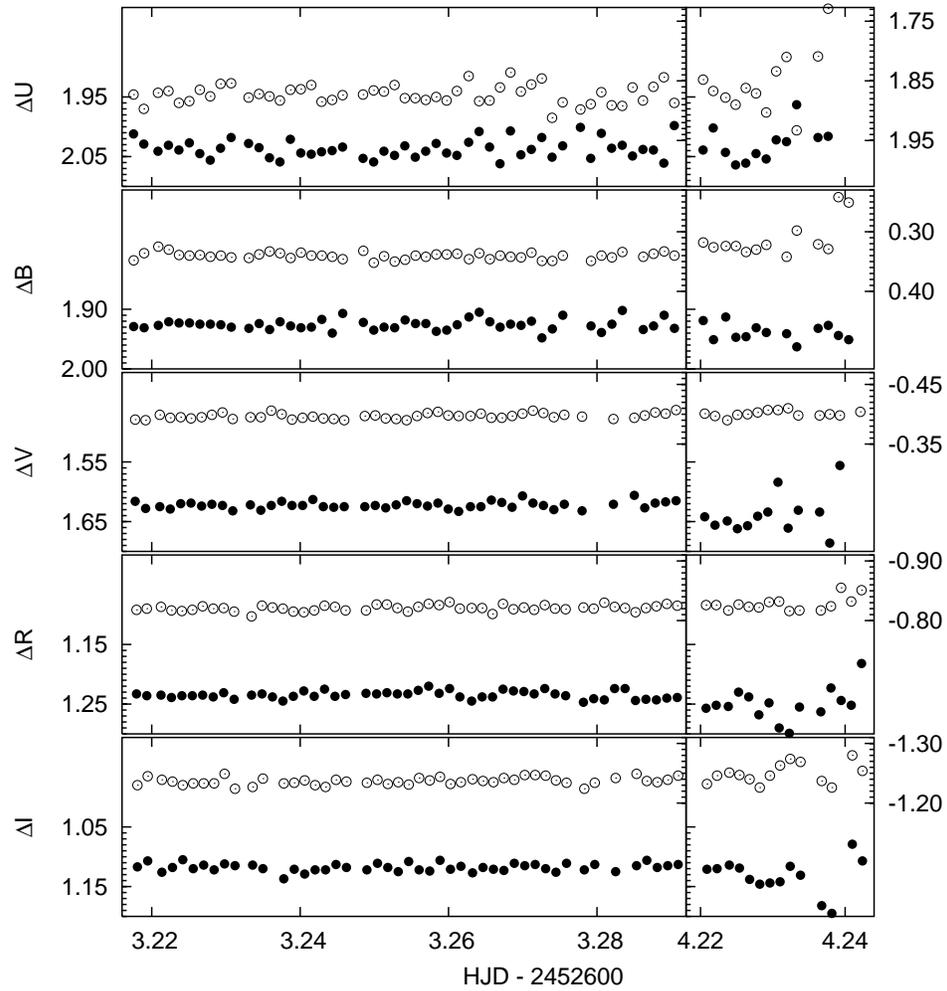}
\caption{The differential light curves $m_{var}$ (dots) and
$m_{comp}$ (open circles) for V471~Per on 24 and 25 November 2002.}
\end{figure}

\newpage
\begin{figure}
\includegraphics[width=10cm]{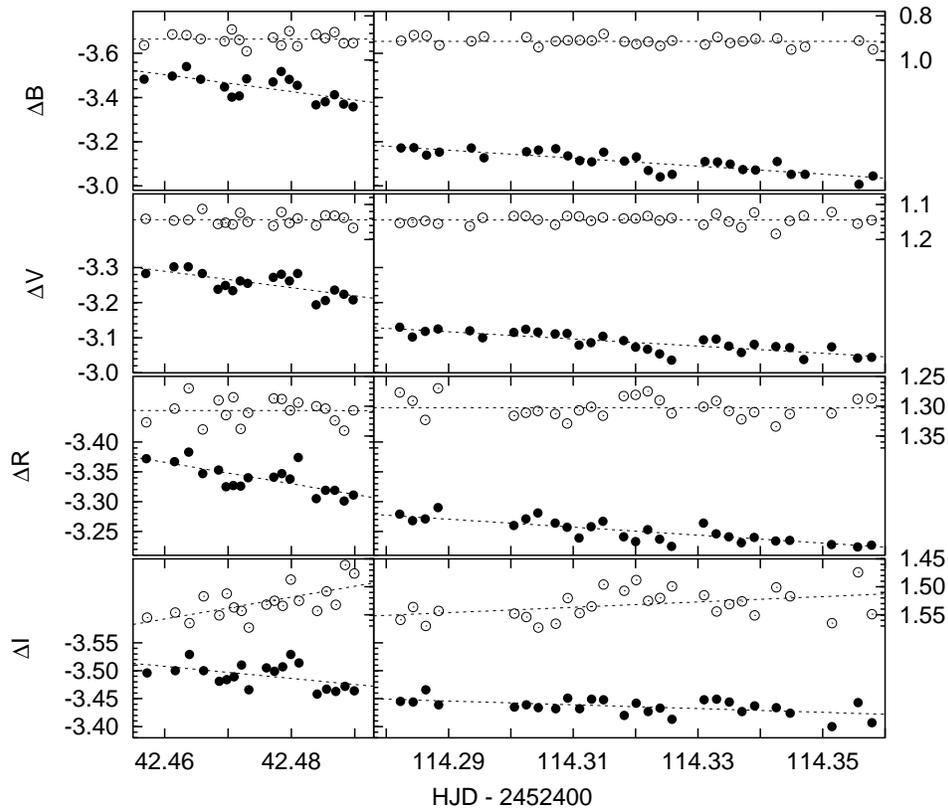}
\caption{Differential light curves  $m_{var}$ (dots) and
$m_{comp}$ (open circles) for RS~Oph obtained on 16 June and 27 August 2002.  
Dashed lines show the linear trends.}
\end{figure}

\newpage
\begin{figure}
\includegraphics[width=10cm]{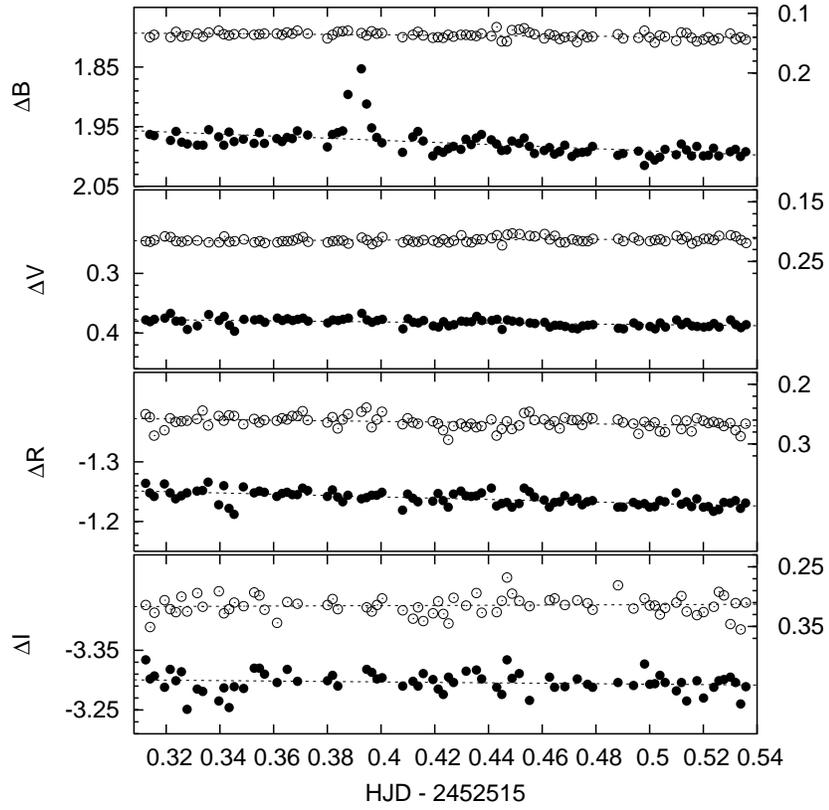}
\caption{Differential light curves $m_{var}$ (dots) and
$m_{comp}$ (open circles) for V627 Cas obtained on 28 August 2002. The flickering-like flare is seen only in {\it B}. 
Dashed lines show the linear trends.}
\end{figure}

\newpage
\begin{figure}
\includegraphics[width=10cm]{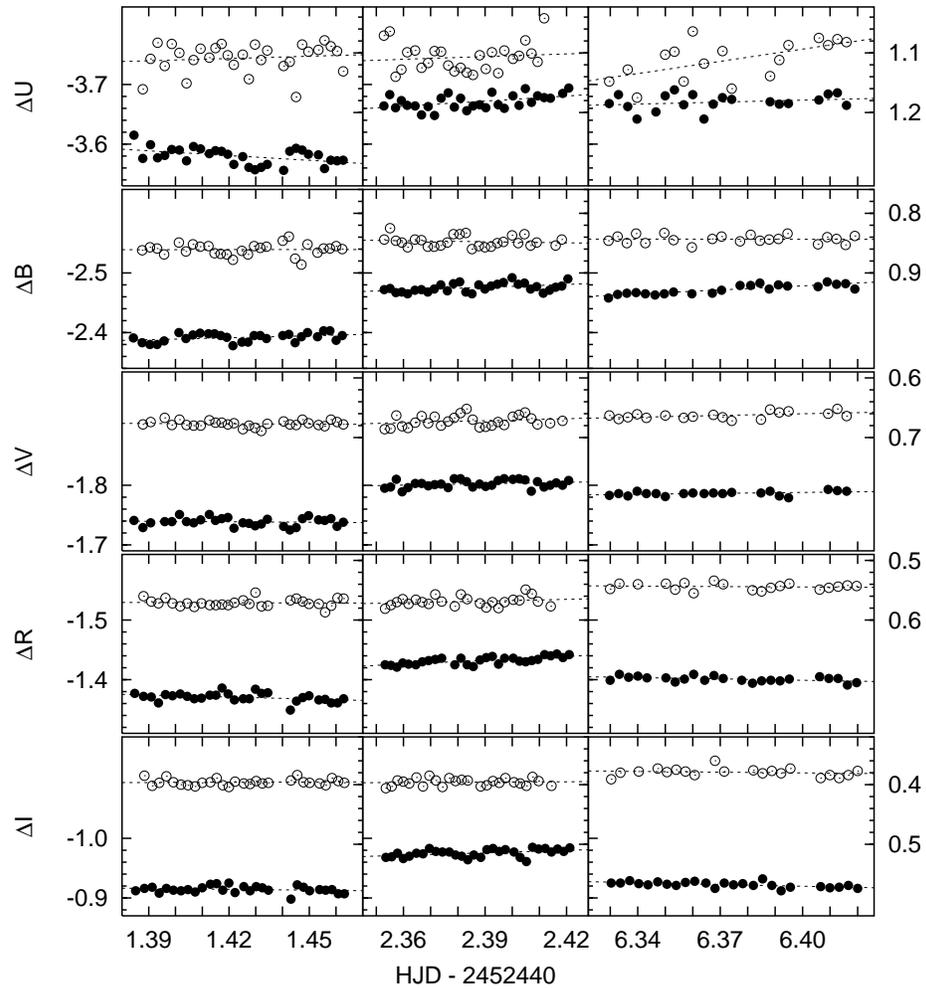}
\caption{The differential light curves $m_{var}$ (dots) and
$m_{comp}$ (open circles) for V886~Her on 15,16 and 20 June 2002. Dashed lines show the linear trends.}
\end{figure}

\newpage
\begin{figure}
\includegraphics[width=10cm]{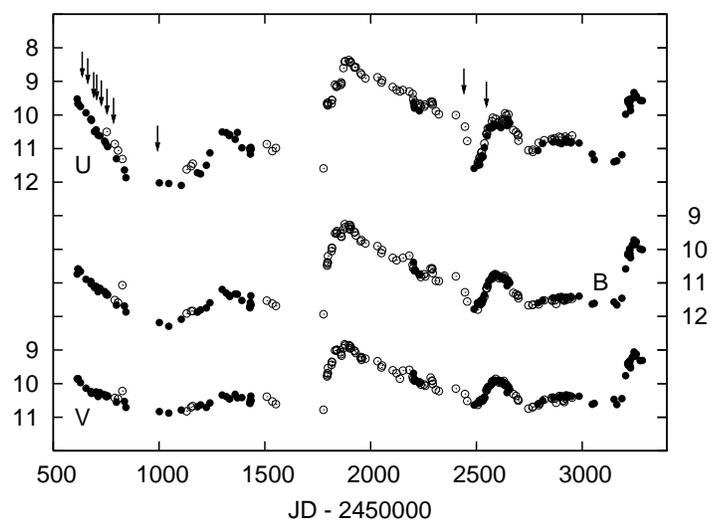}
\caption{The UBV light curves of Z And during 1997-2004 collected by  
Skopal \etal (2002,2004) (open circles) and by Miko{\l}ajewski \& Ga{\l}an (in preparation) 
at Torun Observatory (dots). The arrows on the left indicate the moments of Sokoloski \& Bildsten (1999) {\it B} 
observations. The arrow in minimum $\sim$JD~24501000 indicates the absence of rapid variations. 
The two last arrows around JD~24502500 mark the dates of our {\it UBVRI} data.}
\end{figure}

\newpage
\begin{figure}
\includegraphics[width=10cm]{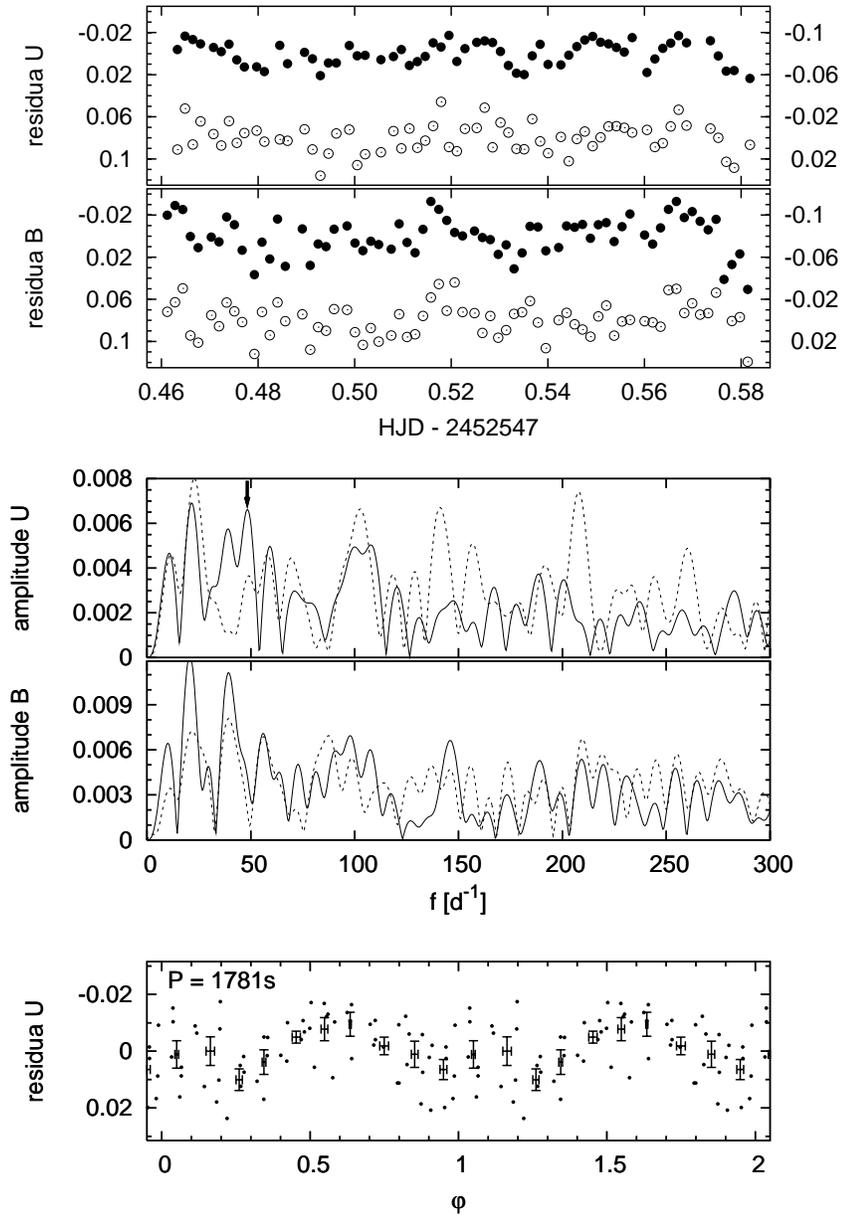}
\caption{The 29 minutes oscillations in Z And on 29 September 2002. 
Top: {\it U} and {\it B} instrumental magnitude residuals of Z~And (dots and left scale) and of 
the brighter comparison star (open circles and right scale).  
Middle: Power spectra of the {\it UB} residuals for Z And (solid line) and the  comparison star (dashed line). 
The arrow points the frequency corresponding to the 29 minutes period.
Bottom: {\it U} residuals folded with this period in 0.1 phase bins, with
corresponding error bars and individual observations (points).}
\end{figure}

\newpage
\begin{figure}
\includegraphics[width=10cm]{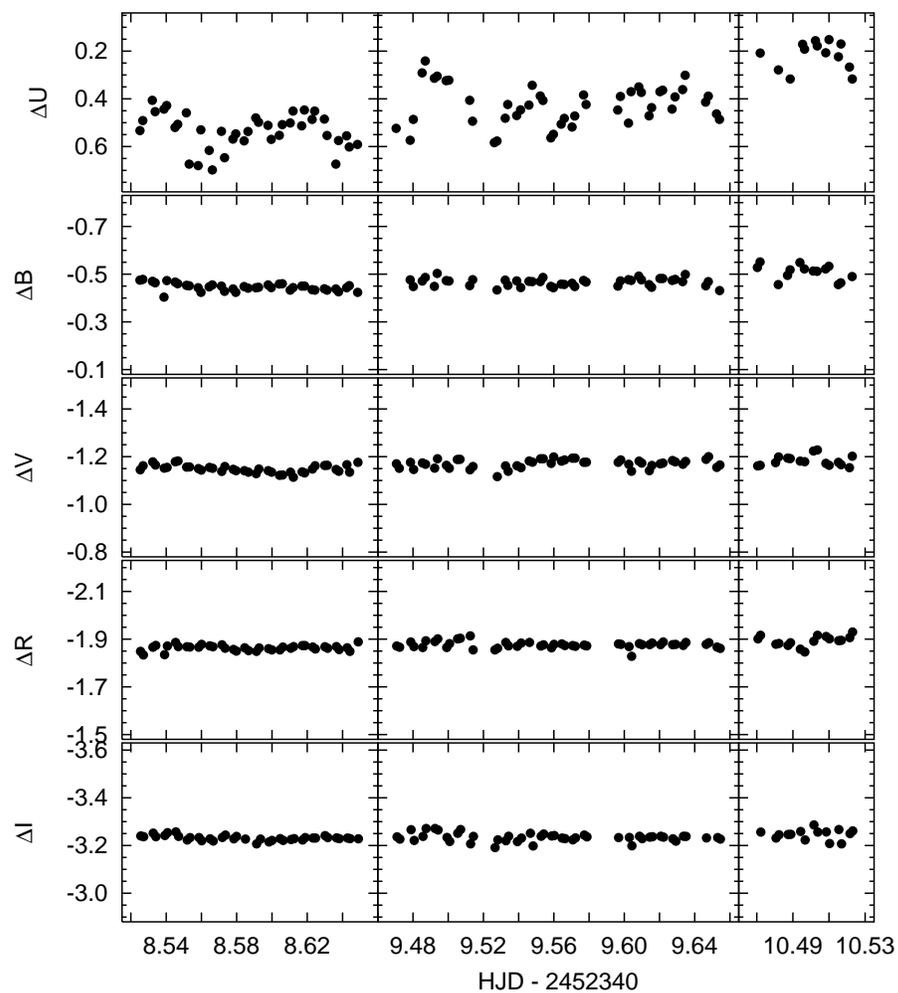}
\caption{The differential light curves $m_{var}$  
for  T~CrB on 14, 15 and 16 March 2002. Significant brightness variations are 
seen only in {\it U}.}
\end{figure}

\newpage
\begin{figure}
\includegraphics[width=10cm]{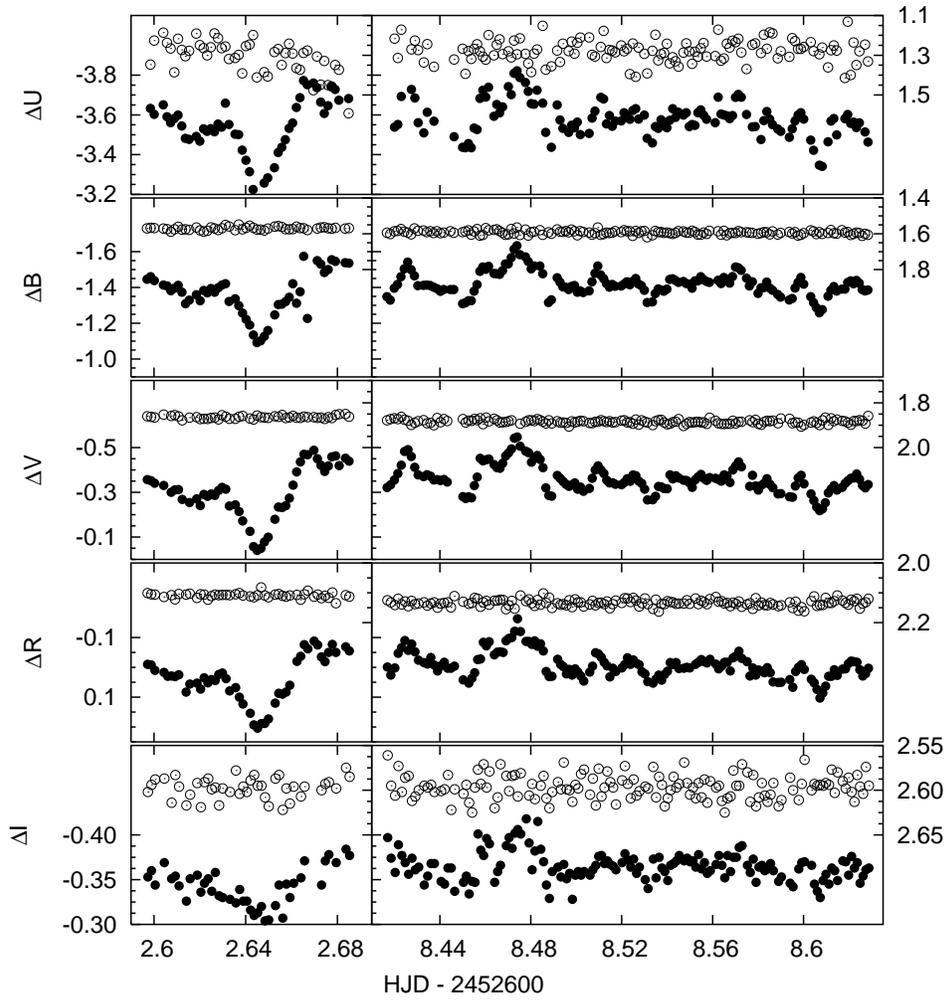}
\caption{Differential light curves $m_{var}$ (dots) and
$m_{comp}$ (open circles) for  MWC~560 on 23 and 29 November 2002.}
\end{figure}

\newpage
\begin{figure}
\includegraphics[width=10cm]{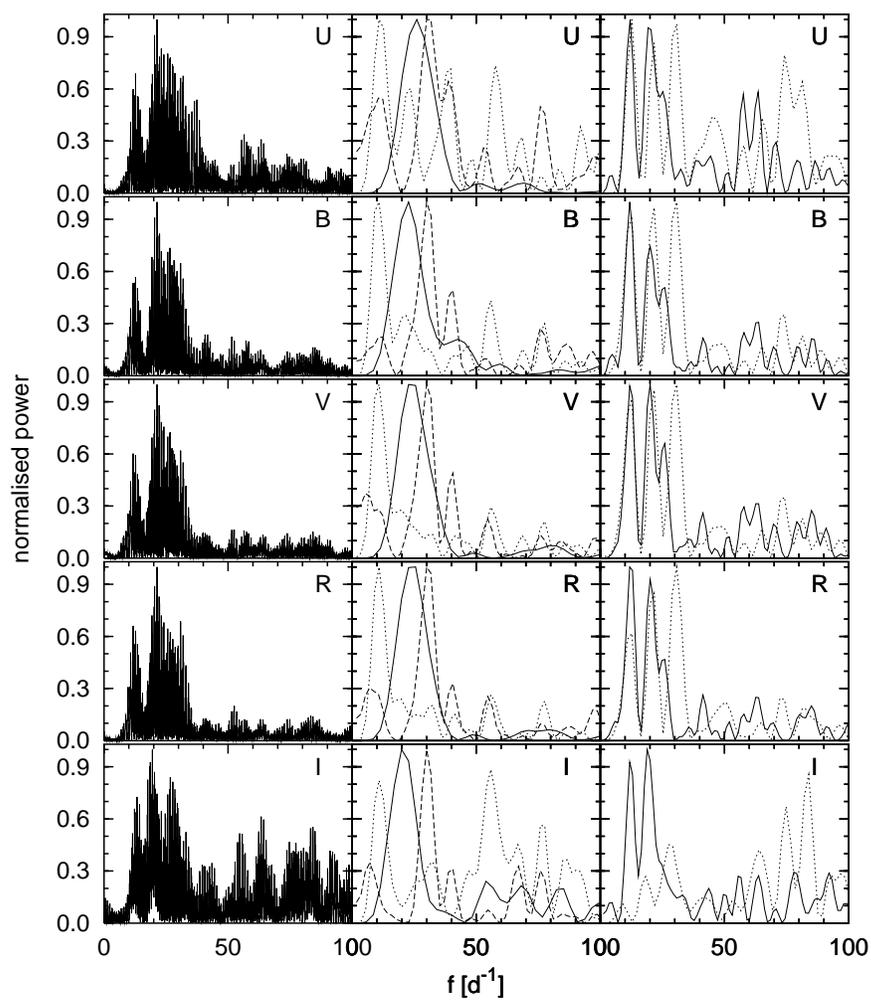}
\caption{Normalized power spectra of the differential light curves of MWC 560.
Left panel: power spectra of all the data obtained in November 2002. The strong peaks 
at frequencies 11.64, 20.37 and 30.14 $d^{-1}$ correspond to periods $\sim$2 hours, $\sim$70, 
$\sim$50 minutes, respectively. Middle panel: power spectra of data obtained on 23 (solid), 
24 (dashed), and 25 (dotted) November 2002. Right panel: power spectra of data obtained on 29 
(solid) and 30 (dotted) November 2002.}
\end{figure}

\newpage
\begin{figure}
\includegraphics[width=10cm]{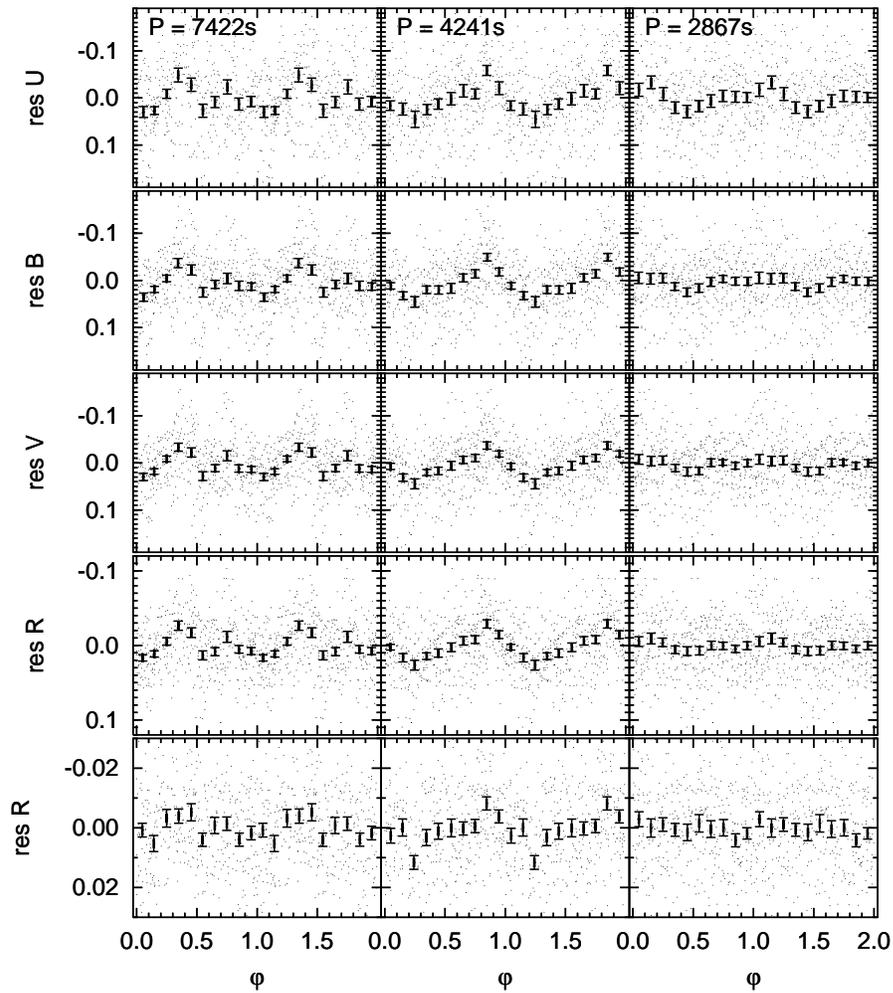}
\caption{MWC 560 differential light curves folded with
the periods found: 7422 sec ($\sim 124$ min), 4241 sec ($\sim 70$ min),
2867 sec ($\sim 48$ min). The November 2002 data were also averaged 
into 0.1 phase bins. The bars show the rms of averaged magnitude in each bin.}
\end{figure}

\newpage
\begin{figure}
\includegraphics[width=10cm]{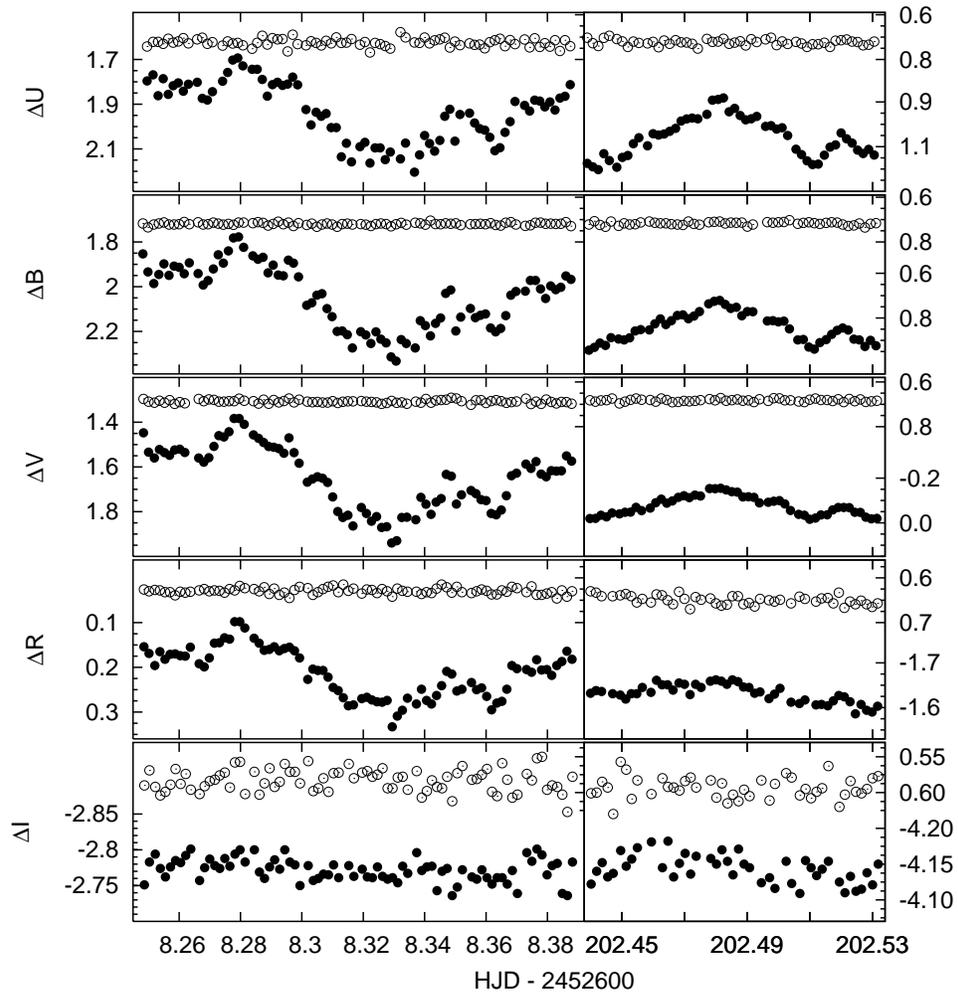}
\caption{Differential light curves $m_{var}$ (dots) and
$m_{comp}$ (open circles) for  V407~Cyg on 29 November 2002 and 11 June 2003. 
In June 2003 the Mira companion was in maximum and V407 Cyg was much brighter 
(the bottom magnitude scale on the right).}
\end{figure}

\newpage
\begin{figure}
\includegraphics[width=10cm]{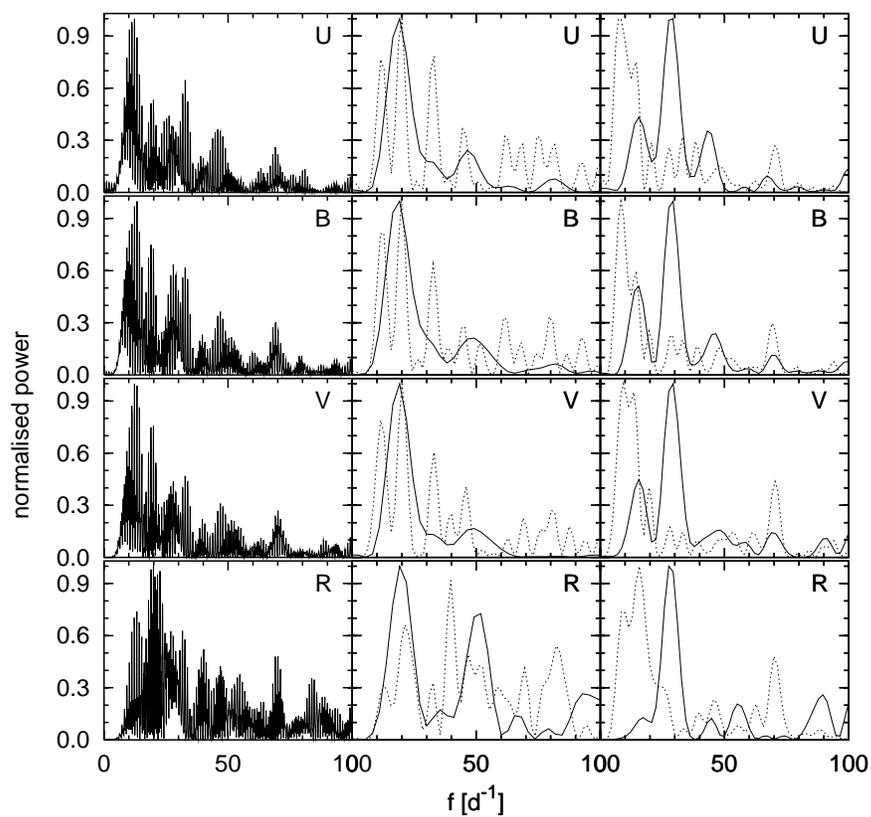}
\caption{Normalized power spectra of V407 Cyg differential light curves.
Left panel: power spectra of combined data set of the four consecutive nights in June 2003. 
The strong peaks 
at frequencies 12, 20 and 32$d^{-1}$ correspond to periods $\sim$2 hours, $\sim$72, 
$\sim$45 minutes, respectively.
Middle panel: power spectra of data obtained on 11 (solid) and 12 (dotted) June 2003. 
Right panel: power spectra of data obtained on 13 (solid) and 14 (dotted) June 2003.}
\end{figure}

\end{document}